\documentclass[12pt]{article}
\usepackage{amssymb}
\usepackage{amsmath}
\usepackage{amsthm}
\usepackage{multirow}

\usepackage[utf8]{inputenc}

\usepackage[normalem]{ulem}

\allowdisplaybreaks

\numberwithin{equation}{section}
\newcommand{\be}{\begin{eqnarray}}
\newcommand{\ee}{\end{eqnarray}}
\newcommand{\non}{\nonumber}
\newcommand{\id}{\mathbb{I}}
\newcommand{\C}{\mathbb{C}}
\newcommand{\tr}{\mathop{\rm tr}\nolimits}

\newcommand{\diag}{\mathop{\rm diag}\nolimits}

\newcommand{\mA}{\mathcal{A}}
\newcommand{\mB}{\mathcal{B}}
\newcommand{\mC}{\mathcal{C}}
\newcommand{\mD}{\mathcal{D}}

\newcommand{\mF}{\mathcal{F}}
\newcommand{\mE}{\mathcal{E}}
\newcommand{\mH}{\mathcal{H}}

\newcommand{\mQ}{\mathcal{Q}}

\newcommand{\refs}{|0\rangle}
\newcommand{\drefs}{\langle0|}

\usepackage[usenames,dvipsnames]{color}

\newtheorem*{prop*}{Proposition}

\begin{document}

\begin{titlepage}
\strut\hfill UMTG--293
\vspace{.5in}
\begin{center}

\LARGE 
A tale of two Bethe 
ans\"atze\\
\vspace{1in}
\large
Antonio Lima-Santos \footnote{
Departamento de F\'{i}sica, Universidade Federal de S\~{a}o Carlos, Caixa Postal 676, 
CEP 13565-905, S\~{a}o Carlos, Brazil, dals@df.ufscar.br},
Rafael I. Nepomechie \footnote{Physics Department,
P.O. Box 248046, University of Miami, Coral Gables, FL 33124 USA, nepomechie@miami.edu}
and Rodrigo A. Pimenta \footnote{Instituto de F\'{i}sica de S\~{a}o Carlos, Universidade de S\~{a}o Paulo, Caixa
Postal 369, 13566-590, S\~{a}o Carlos, SP, Brazil, pimenta@ifsc.usp.br}
\\[0.8in]
\end{center}

\vspace{.5in}

\begin{abstract}
We revisit the construction of the eigenvectors of the single and
double-row transfer matrices associated with the Zamolodchikov-Fateev
model, within the algebraic Bethe ansatz method.  The left and right
eigenvectors are constructed using two different methods: the fusion
technique and Tarasov's construction.  A simple
explicit relation between the eigenvectors from the
two Bethe ans\"atze is obtained.  As a consequence, we obtain the
Slavnov formula for the scalar product between on-shell and off-shell
Tarasov-Bethe vectors.
\end{abstract}

\end{titlepage}

\setcounter{footnote}{0}

\section{Introduction}
The Zamolodchikov-Fateev (ZF) model \cite{Zamolodchikov:1980ku} is a
nineteen-vertex model that can be seen as a 
generalization of the six-vertex model.  Indeed, while the latter is
associated with the spin-$\frac{1}{2}$ XXZ chain, the former
describes an integrable spin-1 XXZ chain.
The ZF R-matrix is a solution of the Yang-Baxter equation, 
which is given by a $9\times
9$ matrix with nineteen non-null entries.  Starting from this 
R-matrix, one can use the quantum
inverse scattering method
\cite{Faddeev:1979gh,Faddeev:1996iy,Sklyanin:1988yz} to 
construct single and double-row commuting transfer
matrices, which are associated with closed and open spin chains,
respectively.

The spectral problem for the transfer matrix can be solved by means of
the algebraic Bethe ansatz.  For the ZF model, the construction of the
Bethe vectors of the single and double-row transfer matrices can be
performed in two different ways.  The first one is by means of the
fusion technique \cite{Kulish:1981gi,
Kulish:1981bi,Babujian:1983ae,A.:1982zz,
Kulish:1983md,SOGO198451,Babujian:1986ua,Kirillov1986,Gohmann:2010se,
Beisert:2015msa,Mezincescu:1990fc,Mezincescu:1991ke,
Nepomechie:2015zwa}, where the auxiliary space is still 
2-dimensional, and therefore 
the Bethe vectors are obtained
similarly to the six-vertex model case. We shall refer to these as 
fusion-Bethe vectors.
The second way is by means of
Tarasov's construction \cite{Tarasov}, which 
entails working with a 3-dimensional auxiliary space. This approach was
originally developed for
the Izergin-Korepin model \cite{izergin1981}\footnote{This model is
also a nineteen-vertex model, 
but it
cannot be obtained by fusion from the fundamental six-vertex model. 
See, however, \cite{Jones:2017}.}, 
see also \cite{Fan:1997wpk} for the double-row case, and was applied 
to the ZF model in
\cite{LimaSantos:1998te,Kurak:2004ip}. We shall refer to these as 
Tarasov-Bethe vectors.

The scalar product between on-shell and off-shell fusion-Bethe
vectors, \textit{i.e.}, a formula of the type found in the celebrated
paper by Slavnov \cite{Sla89}, has been obtained for closed chains,
see \cite{Kitanine:2001bq} for the rational case and
\cite{Deguchi:2009zz} for the trigonometric case.  For open chains,
the Slavnov formula has been obtained only for spin $\frac{1}{2}$, see
\cite{Wang2002633,Belliard:2016ist} for the rational case and
\cite{Kitanine:2007bi} for the trigonometric case.  On the other hand,
Slavnov-type formulas for Tarasov-Bethe vectors have not yet been
found, to the best of our knowledge.  This is probably due to the
intricacy of the exchange relations arising from the Yang-Baxter and
reflection algebras when the auxiliary space has dimension greater
than two, as is the case in Tarasov's construction.

The purpose of this note is to initiate the investigation of the
Slavnov scalar products for Tarasov-Bethe vectors.  Here we handle
this problem by showing that the fusion-Bethe vectors 
and the Tarasov-Bethe vectors are simply related: the result for the
closed chain is given by
(\ref{relationSR},\ref{drelationSR}) and for the open chain\footnote{For simplicity, we focus here on the case of diagonal
boundary conditions.} it is given by
(\ref{relationDR},\ref{drelationDR}).  Thanks to these simple
relations, we can easily obtain the Slavnov formula for the
Tarasov-Bethe vectors, for both the closed (\ref{SLAVBESRZF}) and
open chains (\ref{SLAVBEDRZF}), without going through
Yang-Baxter and reflection algebras in Tarasov's construction.

This paper is organized as follows.  In Section \ref{sec:Fusion} we
recall the basics of the fusion technique.  The fundamental and fused
monodromy and transfer matrices are given in Section
\ref{sec:transfer}.  In Section \ref{sec:single} the Bethe ansatz is
implemented and the Slavnov formula is obtained for the single-row
transfer matrix, while in Section \ref{sec:double} these
results are generalized to the double-row transfer matrix.  Our
concluding remarks are given is Section \ref{sec:conclusion}.

\section{Fusion for R-matrices and K-matrices}\label{sec:Fusion}

In this section we recall some basic ingredients of the fusion
technique, by means of which we can construct integrable
generalizations of the fundamental six-vertex model, for both
single-row \cite{Kulish:1981gi,
Kulish:1981bi,Babujian:1983ae,A.:1982zz,Kulish:1983md,SOGO198451,Babujian:1986ua,Kirillov1986,Gohmann:2010se,
Beisert:2015msa} and double-row
\cite{Mezincescu:1990fc,Mezincescu:1991ke,Nepomechie:2015zwa} transfer
matrices.

\subsection{R-matrices}\label{sec:Rmat}

We start with the fundamental R-matrix
\be
R^{(\tfrac{1}{2},\tfrac{1}{2})}(u)=\left(\begin{array}{cccc}
\sinh(u+\eta) & 0 & 0 & 0\\
0 & \sinh(u) & \sinh(\eta) & 0 \\
0 & \sinh(\eta) & \sinh(u) & 0 \\
0 & 0 & 0 & \sinh(u+\eta)
\end{array}\right)\,,
\label{R11}
\ee
which acts on $\C^{2} \otimes \C^{2}$ and where $\eta$ is the 
free anisotropy parameter. This R-matrix has the parity symmetry
\be
R(u) = {\cal P}\, R(u)\,  {\cal P} \,,
\label{parity}
\ee 
where ${\cal P}$ is the permutation matrix, and the unitarity property
\be
R(u)\, R(-u) = \xi(u)\, \xi(-u) \id\otimes\id\,,
\label{unitarity}
\ee
with $\xi(u) =  \xi^{(\tfrac{1}{2},\tfrac{1}{2})}(u) = \sinh(u+\eta)$.
It also has crossing symmetry
\be
R_{12}(u)=V_1\, R_{12}^{t_2}(-u-\rho)\, V_1
= V_2^{t_2}\, R_{12}^{t_1}(-u-\rho)\, V_2^{t_2} \,,
\label{crossing}
\ee
with 
\be
V=V^{(\tfrac{1}{2},\tfrac{1}{2})} = \left(
\begin{array}{cc}
0 & 1 \\
1 & 0
\end{array} \right)\,, \qquad \rho=\rho^{(\tfrac{1}{2},\tfrac{1}{2})} =\eta + i \pi \,.
\ee
Hence, the matrix $M$ defined by
\be
M = V^{t}\, V 
\label{Mmat}
\ee
is given by 
\be
M = M^{(\tfrac{1}{2},\tfrac{1}{2})} = \id.
\ee

By means of the fusion procedure, we can construct R-matrices $R^{(i,j)}(u)$
associated with $su(2)$ representations of spins $i$ and $j$, with $i, 
j \in \{ \tfrac{1}{2}, 1 \}$. 
These R-matrices,  which map 
$\C^{2i+1} \otimes \C^{2j+1} \mapsto \C^{2i+1} \otimes \C^{2j+1}$,  
satisfy the various Yang-Baxter equations
\be
R_{12}^{(i,j)}(u - v)\,  R_{13}^{(i,k)}(u)\, R_{23}^{(j,k)}(v) 
= R_{23}^{(j,k)}(v)\,  R_{13}^{(i,k)}(u)\, R_{12}^{(i,j)}(u - v)
\,.  \label{YBE}
\ee
Using the fusion procedure (following \cite{Gohmann:2010se,Beisert:2015msa}), we define the 
fused R-matrix
\be
R^{(\tfrac{1}{2},1)}_{1\langle 23\rangle}(u)
&=&\frac{1}{\sinh(u+\tfrac{\eta}{2})} F_{\langle 23\rangle}\,
R^{(\tfrac{1}{2},\tfrac{1}{2})}_{13}(u+\tfrac{\eta}{2})\,
R^{(\tfrac{1}{2},\tfrac{1}{2})}_{12}(u-\tfrac{\eta}{2})\,
E_{\langle 23\rangle} \non\\
&=&  \left(
\begin{array}{cccccc}
\sinh(u+\tfrac{3\eta}{2}) & 0 & 0 & 0 & 0 & 0 \\
0 & \sinh(u+\tfrac{\eta}{2}) & 0 & \tfrac{1}{\sqrt{2}}\sinh(2\eta) & 0 & 0 \\
0 & 0 & \sinh(u-\tfrac{\eta}{2}) & 0 & \sqrt{2}\sinh(\eta)  & 0 \\
0 & \sqrt{2}\sinh(\eta) & 0  & \sinh(u-\tfrac{\eta}{2}) & 0 & 0 \\
0 & 0 & \tfrac{1}{\sqrt{2}}\sinh(2\eta) & 0 & \sinh(u+\tfrac{\eta}{2}) & 0 \\
0 & 0 & 0 & 0 & 0 & \sinh(u+\tfrac{3\eta}{2})
\end{array}\right) \,, \non \\
\label{R12}
\ee
which acts on $\C^{2} \otimes \C^{3}$. Here
\be
E=\left(\begin{array}{ccc}
1 & 0 & 0 \\
0 & \tfrac{1}{\sqrt{2}} & 0 \\
0 & \tfrac{1}{\sqrt{2}} & 0 \\
0 & 0 & 1
\end{array}\right)\,, \qquad F = E^{t} = \left(
\begin{array}{cccc}
 1 & 0 & 0 & 0 \\
 0 & \frac{1}{\sqrt{2}} & \frac{1}{\sqrt{2}} & 0 \\
 0 & 0 & 0 & 1 \\
\end{array}
\right)\,.
\ee
This R-matrix has the unitarity property (\ref{unitarity}) with
$\xi(u) =  \xi^{(\tfrac{1}{2},1)}(u) = 
\sinh(u+\tfrac{3\eta}{2})$.

We also define the fused R-matrix
\be
R^{(1,\tfrac{1}{2})}_{\langle 12\rangle 3}(u)
=\frac{1}{\sinh(u+\tfrac{\eta}{2})} F_{\langle 12\rangle}\,
R^{(\tfrac{1}{2},\tfrac{1}{2})}_{13}(u+\tfrac{\eta}{2})\,
R^{(\tfrac{1}{2},\tfrac{1}{2})}_{23}(u-\tfrac{\eta}{2})\,
E_{\langle 12\rangle}\,,
\label{R21}
\ee
which acts on $\C^{3} \otimes \C^{2}$. The R-matrices (\ref{R12}) and 
(\ref{R21}) are related by
\be
R^{(1,\tfrac{1}{2})}(u) = {\cal P}^{(\tfrac{1}{2},1)} \, 
R^{(\tfrac{1}{2},1)}(u)\, {\cal P}^{(1,\tfrac{1}{2})}  \,,
\ee
where 
\be
{\cal P}^{(\tfrac{1}{2},1)} = \left(
\begin{array}{cccccc}
1 & 0 & 0 & 0 & 0 & 0 \\
0 & 0 & 0 & 1 & 0 & 0 \\
0 & 1 & 0 & 0 & 0 & 0 \\
0 & 0 & 0 & 0 & 1 & 0 \\
0 & 0 & 1 & 0 & 0 & 0 \\
0 & 0 & 0 & 0 & 0 & 1
\end{array}\right)
\ee
is a permutation matrix that maps $\C^{2} \otimes \C^{3}$  to $\C^{3} 
\otimes \C^{2}$, and ${\cal P}^{(1,\tfrac{1}{2})} = \big({\cal 
P}^{(\tfrac{1}{2},1)}\big)^{-1}$.

Finally, we define the fused R-matrix
\be
R^{(1,1)}_{\langle 12\rangle \langle 34\rangle}(u) &=&
F_{\langle 12 \rangle}\, R^{(\tfrac{1}{2},1)}_{1\langle 
34\rangle}(u+\tfrac{\eta}{2})\,
R^{(\tfrac{1}{2},1)}_{2\langle 34\rangle}(u-\tfrac{\eta}{2})\,
E_{\langle 12 \rangle} \non \\
&=& \left(
\begin{array}{ccccccccc}
 a(u) & 0 & 0 & 0 & 0 & 0 & 0 & 0 & 0 \\
 0 & b(u) & 0 & c(u) & 0 & 0 & 0 & 0 & 0 \\
 0 & 0 & f(u) & 0 & d(u) & 0 & h(u) & 0 & 0 \\
 0 & c(u) & 0 & b(u) & 0 & 0 & 0 & 0 & 0 \\
 0 & 0 & \tilde{d}(u) & 0 & e(u) & 0 & \tilde{d}(u)
   & 0 & 0 \\
 0 & 0 & 0 & 0 & 0 & b(u) & 0 & c(u) & 0 \\
 0 & 0 & h(u) & 0 & d(u) & 0 & f(u) & 0 & 0
   \\
 0 & 0 & 0 & 0 & 0 & c(u) & 0 & b(u) & 0 \\
 0 & 0 & 0 & 0 & 0 & 0 & 0 & 0 & a(u) \\
\end{array}
\right) \,,
\label{R22}
\ee
with
\be
a(u) &=& \sinh(u+\eta)\sinh(u+2\eta),~~~b(u) = 
\sinh(u)\sinh(u+\eta),\non \\
d(u) &=&  2\sinh(u)\sinh(\eta),~~~ \tilde{d}(u) = 
2\sinh(u)\sinh(\eta)\cosh^{2}(\eta), \non \\
c(u) &=& \sinh(2\eta)\sinh(u+\eta),~~~
f(u) = \sinh(u)\sinh(u-\eta), \non\\
e(u) &=& \sinh(u)\sinh(u+\eta)+ \sinh(2\eta)\sinh(\eta),~~
h(u) =  \sinh(2\eta)\sinh(\eta)\,,  \label{wgZF}
\ee
which acts on $\C^{3} \otimes \C^{3}$.
This R-matrix has the parity symmetry (\ref{parity}) and the
unitarity property (\ref{unitarity}) with
$\xi(u) =  \xi^{(1,1)}(u) = 
\sinh(u+2\eta)\sinh(u+\eta)$.
Although it does not have the crossing symmetry 
(\ref{crossing}), it is related to a 
symmetric R-matrix $\tilde R^{(1,1)}(u) =\left[\tilde 
R^{(1,1)}(u)\right]^{t_{1} t_{2}}$ by a constant gauge transformation 
\be
\tilde R^{(1,1)}(u) = C_{1}^{-1}\, C_{2}^{-1}\,  R^{(1,1)}(u)\, C_{1}\, 
C_{2} \,, \qquad C=\diag \left(\cosh(\eta),\cosh(\eta), 1\right) \,;
\label{R22sym}
\ee
and this symmetric R-matrix {\em does} have crossing symmetry, with 
\be
V=V^{(1,1)} = \left(
\begin{array}{ccc}
0 & 0 & 1 \\
0 & 1 & 0 \\
1 & 0 & 0
\end{array} \right)\,, \qquad \rho=\rho^{(1,1)} =\eta + i \pi \,,
\ee
and therefore 
\be
M = M^{(1,1)} = \id.
\ee
The R-matrix (\ref{R22sym}) was firstly obtained by Zamolodchikov and
Fateev \cite{Zamolodchikov:1980ku}.

\subsection{K-matrices}\label{sec:Kmat}

We now construct corresponding K-matrices $K^{\pm\, (i)}(u)$, which 
map $\C^{2i+1} \mapsto \C^{2i+1}$. For $K^{-\, (i)}(u)$, the 
boundary Yang-Baxter equations are 
\be
R_{12}^{(i,j)}(u - v)\, K^{- (i)}_1(u)\ R_{21}^{(j,i)} (u + v)\, K^{- (j)}_2(v)
= K^{- (j)}_2(v)\, R_{12}^{(i,j)}(u + v)\, K^{- (i)}_1(u)\, 
R_{21}^{(j,i)}(u - v)  \,, \non\\
\label{BYBEm}
\ee
where
\be
R_{21}^{(j,i)}(u) = {\cal P}^{(i,j)}\, R_{12}^{(j,i)}(u)\, {\cal 
P}^{(j,i)} \,.
\ee

For the fundamental $K^{-}$-matrix, we restrict for simplicity to the diagonal solution
\be
K^{-\, (\tfrac{1}{2})}(u) = \diag\left(\sinh(u+ \xi^{-})\,,  -\sinh(u- 
\xi^{-}) \right) \,,
\label{Km1}
\ee
where $\xi^{-}$ is an arbitrary boundary parameter. For $K^{+}$, we take
\be
K^{+\, (\tfrac{1}{2})}(u) &=& K^{-\, 
(\tfrac{1}{2})}(-u-\rho^{(\tfrac{1}{2},\tfrac{1}{2})})\Big\vert_{\xi^{-}\rightarrow \xi^{+}} \non\\
&=& 
\diag\left(\sinh(u+ \eta -\xi^{+})\,,  -\sinh(u+\eta+ 
\xi^{+}) \right)\,,
\ee
where $\xi^{+}$ is another arbitrary boundary parameter.

The fused $K^{-}$-matrix is given by (following \cite{Nepomechie:2015zwa})
\be
K^{-\, (1)}(u) &=& \frac{1}{2\sinh(u+\tfrac{\eta}{2})} F_{\langle 12 
\rangle}\, K^{-\, (\tfrac{1}{2})}_{1}(u+\tfrac{\eta}{2})\,
R^{(\tfrac{1}{2},\tfrac{1}{2})}_{12}(2u)\,
K^{-\, (\tfrac{1}{2})}_{2}(u-\tfrac{\eta}{2})\, {\cal P}_{12}\, 
E_{\langle 12 \rangle} \non\\
&=& \diag\left( k^{-}_{1}(u)\,, k^{-}_{2}(u)\,, k^{-}_{3}(u) \right) \,,
\label{Km2}
\ee 
where
\be
k^{-}_{1}(u) &=& \cosh(u+\tfrac{\eta}{2})\, 
\sinh(u+\tfrac{\eta}{2}+\xi^{-})\, \sinh(u-\tfrac{\eta}{2}+\xi^{-}) 
\,, \non\\
k^{-}_{2}(u) &=& -\cosh(u+\tfrac{\eta}{2})\, 
\sinh(u-\tfrac{\eta}{2}+\xi^{-})\, \sinh(u-\tfrac{\eta}{2}-\xi^{-}) 
\,, \non\\
k^{-}_{3}(u) &=& \cosh(u+\tfrac{\eta}{2})\, 
\sinh(u+\tfrac{\eta}{2}-\xi^{-})\, \sinh(u-\tfrac{\eta}{2}-\xi^{-}) 
\,.
\ee
For $K^{+}$, we similarly take
\be
K^{+\, (1)}(u) = K^{-\, 
(1)}(-u-\rho^{(1,1)})\Big\vert_{\xi^{-}\rightarrow \xi^{+}} \,.
\ee

\section{Monodromy and transfer matrices}\label{sec:transfer}

Using the R and K matrices introduced in the previous section, we can obtain
families of commuting operators, \textit{i.e.}, transfer matrices.

\subsection{Single-row}

Indeed, the single-row monodromy matrix for the R-matrix (\ref{R12}) is defined as
\be\label{SRmonodromy12}
T_{a}^{(\tfrac{1}{2},1)}(u) = R_{a N}^{(\tfrac{1}{2},1)}(u) \ldots 
R_{a 1}^{(\tfrac{1}{2},1)}(u) \,, 
\ee
and the associated transfer matrix by
\be\label{SRtransfer12}
t^{\left(\frac{1}{2},1\right)}(u) = \tr_{a} T_{a}^{(\tfrac{1}{2},1)}(u)\,.
\ee
Similarly, for the R-matrix (\ref{R22})
\be\label{SRmonodromy22}
T_{a}^{(1,1)}(u) = R_{a N}^{(1,1)}(u) \ldots R_{a 1}^{(1,1)}(u) \,,
\ee
and
\be\label{SRtransfer22}
t^{\left(1,1\right)}(u) = \tr_{a} T_{a}^{(1,1)}(u)\,.
\ee
The Yang-Baxter equations (\ref{YBE}) imply
\be\label{RTT12}
R_{12}^{(\tfrac{1}{2},\tfrac{1}{2})}(u-v)\, 
T_1^{\left(\frac{1}{2},1\right)}(u)\, 
T_2^{\left(\frac{1}{2},1\right)}(v)=T_2^{\left(\frac{1}{2},1\right)}(v)\, T_1^{\left(\frac{1}{2},1\right)}(u)\, R_{12}^{(\tfrac{1}{2},\tfrac{1}{2})}(u-v)
\ee
and
\be\label{RTT22}
R_{12}^{(1,1)}(u-v)\, T_1^{\left(1,1\right)}(u)\, 
T_2^{\left(1,1\right)}(v)=T_2^{\left(1,1\right)}(v)\, 
T_1^{\left(1,1\right)}(u)\, R_{12}^{(1,1)}(u-v) \,,
\ee
as well as
\be
\left[ t^{(\tfrac{1}{2},1)}(u) \,,  t^{(\tfrac{1}{2},1)}(v) \right] = 
0 \,, \qquad
\left[ t^{(1,1)}(u) \,,  t^{(1,1)}(v) \right] = 
0 \,, 
\ee
and
\be
\left[ t^{(\tfrac{1}{2},1)}(u) \,,  t^{(1,1)}(v) \right] = 0 \,.
\ee
The latter relation implies that $t^{(\tfrac{1}{2},1)}(u)$ and 
$t^{(1,1)}(v)$ can be diagonalized simultaneously. In addition, one 
obtains with the help of (\ref{R22}) the important
relation
\be\label{SRmono22to12}
T_{\langle 12 \rangle}^{(1,1)}(u) =  F_{\langle 12 \rangle}\, 
T_{1}^{(\tfrac{1}{2},1)}(u+\tfrac{\eta}{2})\, 
T_{2}^{(\tfrac{1}{2},1)}(u-\tfrac{\eta}{2})\, E_{\langle 12 \rangle} 
\,,
\ee
which will be used in Section 
\ref{sec:BArelations}.

\subsection{Double-row}

In order to construct double-row objects, one needs to introduce ``reflected'' single-row monodromy matrices
\be
\widehat T_{a}^{(\tfrac{1}{2},1)}(u) = R_{a 1}^{(\tfrac{1}{2},1)}(u) \ldots 
R_{a N}^{(\tfrac{1}{2},1)}(u) \,,
\ee
and similarly
\be
\widehat T_{a}^{(1,1)}(u) = R_{a 1}^{(1,1)}(u) \ldots R_{a 
N}^{(1,1)}(u) \,.
\ee
The corresponding double-row monodromy matrices are then defined as follows
\be
U_{a}^{(\tfrac{1}{2},1)}(u) &=& T_{a}^{(\tfrac{1}{2},1)}(u)\, 
K^{-(\tfrac{1}{2})}_{a}(u)\, \widehat T_{a}^{(\tfrac{1}{2},1)}(u)  
\,, \non \\
U_{a}^{(1,1)}(u) &=& T_{a}^{(1,1)}(u)\, 
K^{-(1)}_{a}(u)\, \widehat T_{a}^{(1,1)}(u) \,.
\ee
They obey the boundary Yang-Baxter equations (\ref{BYBEm}), in 
particular
\be
\lefteqn{R_{12}^{(\tfrac{1}{2}, \tfrac{1}{2})}(u - v)\, 
U^{(\tfrac{1}{2},1)}_1(u)\ R_{12}^{(\tfrac{1}{2},\tfrac{1}{2})} (u + 
v)\, U^{(\tfrac{1}{2},1)}_2(v)} \non\\
&&= U^{(\tfrac{1}{2},1)}_2(v)\, R_{12}^{(\tfrac{1}{2}, \tfrac{1}{2})}(u + v)\, U^{(\tfrac{1}{2},1)}_1(u)\, 
R_{12}^{(\tfrac{1}{2}, \tfrac{1}{2})}(u - v)  \,, 
\label{BYBEU1}
\ee
and
\be
\lefteqn{R_{12}^{(1,1)}(u - v)\, U^{(1,1)}_1(u)\ R_{12}^{(1,1)} (u + 
v)\, U^{(1,1)}_2(v)}\non\\
&&= U^{(1,1)}_2(v)\, R_{12}^{(1,1)}(u + v)\, U^{(1,1)}_1(u)\, 
R_{12}^{(1,1)}(u - v)  \,.
\label{BYBEU2}
\ee
Moreover, they are related by
\be
U^{(1,1)}_{\langle 12 \rangle}(u) &=& \frac{1}{2\sinh(u+\tfrac{\eta}{2})} F_{\langle 12 
\rangle}\, U^{(\tfrac{1}{2},1)}_{1}(u+\tfrac{\eta}{2})\,
R^{(\tfrac{1}{2},\tfrac{1}{2})}_{12}(2u)\,
U^{(\tfrac{1}{2},1)}_{2}(u-\tfrac{\eta}{2})\, {\cal P}_{12}\, 
E_{\langle 12 \rangle} \,, \non\\
\label{Ufusion}
\ee
which can be obtained with the help of (\ref{Km2}). This 
relation, which is the double-row version of (\ref{SRmono22to12}), will 
be exploited in Section \ref{sec:BArelationsDR}.

Finally, the corresponding transfer matrices are given by
\be\label{DRtransfer12}
\tau^{(\tfrac{1}{2},1)}(u) &=& \tr_{a} K^{+(\tfrac{1}{2})}_{a}(u)\, 
U_{a}^{(\tfrac{1}{2},1)}(u) \,,
\ee
\be\label{DRtransfer22}
\tau^{(1,1)}(u) &=& \tr_{a} K^{+(1)}_{a}(u)\, 
U_{a}^{(1,1)}(u) \,.
\ee
One can show that these transfer matrices obey
\be
\left[ \tau^{(\tfrac{1}{2},1)}(u) \,,  \tau^{(\tfrac{1}{2},1)}(v) \right] = 
0 \,, \qquad
\left[ \tau^{(1,1)}(u) \,,  \tau^{(1,1)}(v) \right] = 
0 \,, 
\ee
as well as 
\be
\left[ \tau^{(\tfrac{1}{2},1)}(u) \,,  \tau^{(1,1)}(v) \right] = 0 \,.
\ee
Similarly to the single-row case, the latter relation implies that $\tau^{(\tfrac{1}{2},1)}(u)$ and 
$\tau^{(1,1)}(v)$ can be diagonalized simultaneously.

\section{Bethe ansatz: single-row transfer matrix}\label{sec:single}

In this section we obtain the eigenvectors of the single-row transfer matrices (\ref{SRtransfer12}) and (\ref{SRtransfer22}).

\subsection{Bethe vectors from fusion}

We start by applying the Bethe ansatz to (\ref{SRtransfer12}), and we
refer to
\cite{Kulish:1981gi,Kulish:1981bi,Babujian:1983ae,A.:1982zz,Kulish:1983md,SOGO198451,Babujian:1986ua,Kirillov1986,Gohmann:2010se,
Beisert:2015msa} for more details.  For that, we use the 2-dimensional auxiliary space
representation, \textit{i.e.}, we set
\be\label{rep1}
T^{\left(\frac{1}{2},1\right)}_a(u)=\left(
\begin{array}{cc}
 A(u) & B(u) \\
 C(u) & D(u) 
\end{array}
\right)_a\,,
\ee
in which each entry is an operator acting on the quantum space ${\left(\mathbb{C}^{3}\right)}^{\otimes N}$. These operators
satisfy exchange relations dictated by (\ref{RTT12}).

We also introduce the reference state vector
\be\label{refstate}
\refs=\left( 
\begin{array}{c}
1 \\ 
0 \\
0
\end{array}
\right)^{\otimes N}
\ee
as well as its dual
\be\label{drefstate}
\drefs={\left(
\begin{array}{cccc}1 & 0 & \cdots & 0
\end{array}\right)}^{\otimes N} \,,
\ee
such that $\drefs 0\rangle=1$. The action of the monodromy operators 
on (\ref{refstate},\ref{drefstate}) is given by
\be\label{actionsinglerow}
A(u)\refs = \lambda_1(u) \refs\,,\quad D(u)\refs = \lambda_2(u) \refs\,,\quad C(u)\refs=0\,,
\ee
\be
\drefs A(u) =\drefs \lambda_1(u) \,,\quad \drefs D(u) =\drefs \lambda_2(u) \,,\quad \drefs B(u)=0\,,
\ee
where
\be\label{lambda12}
\lambda_1(u)=\sinh^N\left(u+\frac{3\eta}{2}\right)\,,\quad
\lambda_2(u)=\sinh^N\left(u-\frac{\eta}{2}\right)\,.
\ee
The transfer matrix (\ref{SRtransfer12}) is given by
\be
t^{(\tfrac{1}{2},1)}(u)=A(u)+D(u)\,.
\ee
The Bethe vectors are given by
\be
|\phi_m(u_1,\dots,u_m)\rangle =B(u_1)\dots B(u_m)\refs\,
\label{FBVsingle}
\ee
and
\be
\langle\phi_m(u_1,\dots,u_m)|=\drefs C(u_1)\dots C(u_m)\,.
\label{dFBVsingle}
\ee
In fact, using the Yang-Baxter algebra (\ref{RTT12}) one can show that
\be
t^{\left(\frac{1}{2},1\right)}(u)|\phi_m(u_1,\dots,u_m)\rangle&=&
\lambda(u,u_1,\dots u_m)|\phi_m(u_1,\dots,u_m)\rangle\nonumber\\&+&
\sum_{j=1}^m 
\sinh(\eta)F(u,u_j)\frac{E_j(u_j)}{Q_j(u_j)}B(u)|\phi_{m-1}(u_1,\ldots,\hat{u}_j,\ldots,u_m)\rangle\non\\ 
\label{offshellsingleF}
\ee
and
\be
\langle\phi_m(u_1,\dots,u_m)| t^{\left(\frac{1}{2},1\right)}(u) &=&
\langle\phi_m(u_1,\dots,u_m)| \lambda(u,u_1,\dots u_m)\nonumber\\&+&
\sum_{j=1}^m \langle\phi_m (u_1,\ldots,\hat{u}_j,\ldots,u_m)|C(u) 
\frac{E_j(u_j)}{Q_j(u_j)}F(u,u_j)\sinh(\eta) \,, \non\\
\label{doffshellsingleF}
\ee
where
\be
\lambda(u,u_1,\dots u_m)=\lambda_1(u)\frac{Q(u-\eta)}{Q(u)}+\lambda_2(u)\frac{Q(u+\eta)}{Q(u)}\,,
\label{eigenvalsingleF}
\ee
\be\label{BPsingleF}
E_j(u)=
\lambda_1(u)Q_j(u-\eta)-\lambda_2(u)Q_j(u+\eta)
\,,
\ee
and
\be
F(u,v)=\frac{1}{\sinh(u-v)}\,.
\ee
In the above formulae we have introduced the Baxter Q-polynomial
\be\label{BaxterQ}
Q(u)=\prod_{i=1}^m \sinh(u-u_i)\,,
\ee
which is indeed a Laurent polynomial in $z\equiv e^{u}$,
as well as the indexed Baxter Q-polynomial
\be\label{BaxterQj}
Q_j(u)=\prod_{i\neq j}^m \sinh(u-u_i)\,.
\ee
Let us also introduce the two-index polynomial which we will use later
\be\label{BaxterQjk}
Q_{j,k}(u)=\prod_{i\neq j,k}^m \sinh(u-u_i)\,.
\ee
The equations
\be\label{periodicBEfusion}
E_j(u_j)=0\quad \textrm{for}\quad j=1,\dots,m\,,
\ee
are the conditions for the ``unwanted'' terms in the off-shell 
equations (\ref{offshellsingleF}) and (\ref{doffshellsingleF}) to 
vanish; and therefore, for the Bethe vectors (\ref{FBVsingle}) and 
(\ref{dFBVsingle}) to be eigenvectors of the transfer matrix 
$t^{\left(\frac{1}{2},1\right)}(u)$, with 
corresponding eigenvalue $\lambda(u,u_1,\dots u_m)$ given by (\ref{eigenvalsingleF}).
These polynomial equations are known as the Bethe ansatz equations, and 
we shall refer to $E_j(u)$ as the Bethe ansatz polynomials. As usual, the notation $\hat{u}_j$ means that the
rapidity $u_j$ is absent from a given function or operator argument.

Finally, let us suppose that $\{u_1,\dots,u_m\}$ satisfy the Bethe
equations (\ref{periodicBEfusion}) (\textit{i.e.}, they are on-shell
rapidities) and that there is no restriction on $\{v_1,\dots,v_m\}$
(\textit{i.e.}, they are off-shell rapidities).  Then, the scalar
product $\langle\phi_m(u_1,\dots,u_m)|\phi_m(v_1,\dots,v_m)\rangle$ is
given by the determinant formula \cite{Sla89,Deguchi:2009zz}
\be\label{SLAVSR}
\langle\phi_m(u_1,\dots,u_m)|\phi_m(v_1,\dots,v_m)\rangle=
\prod_{i=1}^m\lambda_2(u_i)
\frac{\textrm{det}_m\left(\frac{\partial}{\partial u_i}\lambda(v_j,u_1,\dots,u_m)\right)}
{\textrm{det}_m\left(F(v_i,u_j)\right)}\,,\non\\
\ee
which is known as the Slavnov formula. Using
\be
\frac{\partial}{\partial 
u_i}\lambda(v_j,u_1,\dots,u_m)=\sinh(\eta)F(u_i,v_j)F(v_j,u_i)\frac{E_i(v_j)}{Q_i(v_j)} \,,
\ee
we can rewrite the Slavnov formula as
\be\label{SLAVBESR}
\langle\phi_m(u_1,\dots,u_m)|\phi_m(v_1,\dots,v_m)\rangle=
\prod_{i=1}^m\lambda_2(u_i)
\frac{\textrm{det}_m\left(\sinh(\eta)F(u_i,v_j)F(v_j,u_i)\frac{E_i(v_j)}{Q_i(v_j)}\right)}
{\textrm{det}_m\left(F(v_i,u_j)\right)}\,. \non\\
\ee

Although the Slavnov formula is frequently written in the form
(\ref{SLAVSR}), which was introduced in \cite{Kitanine:1999rfm}, it
seems more suitable for generalization to 
rewrite the Slavnov formula 
as in (\ref{SLAVBESR}) in
terms of the Bethe ansatz and Baxter polynomials instead of 
the derivative
of the transfer matrix eigenvalue, as it will become clear in 
Section \ref{sec:SPT}.  
The form (\ref{SLAVBESR}) is closer to the 
expression presented by Slavnov
in his pioneering paper \cite{Sla89}.

\subsection{Bethe vectors from Tarasov's construction}
We now apply the Bethe ansatz to (\ref{SRtransfer22}), see \cite{Tarasov,LimaSantos:1998te}. Here, we use the 3-dimensional auxiliary
space representation, \textit{i.e.}, we set
\be\label{rep2}
T^{\left(1,1\right)}_a(u)=\left(
\begin{array}{ccc}
A_1(u) & B_1(u) & B_2(u)\\
C_1(u) & A_2(u) & B_3(u)\\
C_2(u) & C_3(u) & A_3(u)
\end{array}
\right)_a\,,
\ee
where each entry acts on the quantum space ${\left(\mathbb{C}^{3}\right)}^{\otimes N}$. These operators
satisfy exchange relations dictated by (\ref{RTT22}). The action of the monodromy operators (\ref{rep2}) on the reference state (\ref{refstate},\ref{drefstate}) is given by
\be\label{actionref1}
A_j(u)\refs = \Lambda_j(u)\refs\,,\quad C_j(u)\refs = 0\,,
\ee
\be\label{dactionref1}
\drefs A_j(u)=\drefs \Lambda_j(u)\,,\quad \drefs B_j(u)=0\,,
\ee
for $j=1,2,3$ and where
\begin{align}
\Lambda_1(u) &=\left(\sinh (u+\eta) \sinh (u+2 \eta)\right)^N\,,\quad \Lambda_2(u)=\left(\sinh (u) \sinh (u+\eta )\right)^N\,,\non\\
\Lambda_3(u) &=\left(\sinh (u) \sinh (u-\eta )\right)^N\,.
\label{Lambda123}
\end{align}
The transfer matrix (\ref{SRtransfer22}) is given by
\be
t^{\left(1,1\right)}(u)=A_1(u)+A_2(u)+A_3(u)\,.
\ee
The Bethe vector is constructed by means of the recursion relation
\be\label{TBVsingle}
|\psi_m(u_1,\dots,u_m)\rangle&=&B_1(u_1)|\psi_{m-1}(u_2,\ldots,u_m)\rangle\nonumber\\&-&
 B_2(u_1)\sum_{i=2}^m \gamma_i^{(m)}(u_1,\ldots,u_m) 
 |\psi_{m-2}(u_2,\ldots,\hat{u}_i,\ldots,u_m)\rangle \,,
\ee
where
\be\label{defgamma}
\gamma_i^{(m)}(u_1,\ldots,u_m)=2\sinh(\eta)\Lambda_1(u_i)F(u_1,u_i+\eta) \frac{Q_{1,i}(u_i-2\eta)}{Q_{1,i}(u_i)}\prod_{j=2,j<i}^m \Omega(u_i,u_j)\,,
\ee
and
\be\label{defOmega}
\Omega(u,v)=\frac{\sinh (u-v-\eta ) \sinh (u-v+2 \eta )}{\sinh (u-v-2 \eta ) \sinh
   (u-v+\eta )}\,.
\ee
The dual Bethe vector is given by
\be\label{dTBVsingle}
\langle\psi_m(u_1,\dots,u_m)|&=&\langle\psi_{m-1}(u_2,\ldots,u_m)| C_1(u_1)\nonumber\\&-&
\sum_{i=2}^m \tilde\gamma_i^{(m)}(u_1,\ldots,u_m) 
\langle\psi_{m-2}(u_2,\ldots,\hat{u}_i,\ldots,u_m)|  C_2(u_1)\,,
\ee
where
\be
\tilde\gamma_i^{(m)}(u_1,\ldots,u_m)=\sinh(2\eta)\cosh(\eta)\Lambda_1(u_i)F(u_1,u_i+\eta) \frac{Q_{1,i}(u_i-2\eta)}{Q_{1,i}(u_i)} \prod_{j=2,j<i}^m \Omega(u_i,u_j)\,.
\ee
Note that the the initial conditions
\be
|\psi_{0}\rangle = \refs\,,\quad \langle\psi_{0}|=\drefs \,,
\ee
are assumed for (\ref{TBVsingle}) and (\ref{dTBVsingle}).

Using the Yang-Baxter algebra (\ref{RTT22}) one can show that the Tarasov-Bethe vectors (\ref{TBVsingle}) and
(\ref{dTBVsingle}) satisfy the off-shell equations
\be\label{offshellTBV}
&& t^{\left(1,1\right)}(u)|\psi_m(u_1,\ldots,u_m)\rangle =
\Lambda(u,u_1,\ldots,u_m)|\psi_m(u_1,\ldots,u_m)\rangle\nonumber\\&&+
\sum_{j=1}^m \sinh(2\eta) F(u,u_j) \frac{\bar E_j(u_j)Q_j(u_j-2\eta)}{Q_j(u_j)Q_j(u_j-\eta)}\prod_{p<j}^m\Omega(u_j,u_p)B_1(u)|\psi_{m-1}(u_1,\ldots,\hat{u}_j,\ldots,u_m)\rangle\nonumber\\
&&+
\sum_{j=1}^m 2\sinh(\eta)F(u,u_j+\eta) \frac{\bar E_j(u_j)Q_j(u_j-2\eta)}{Q_j(u_j)Q_j(u_j-\eta)}\prod_{p<j}^m\Omega(u_j,u_p)B_3(u)|\psi_{m-1}(u_1,\ldots,\hat{u}_j,\ldots,u_m)\rangle\nonumber\\
&&+
\sum_{j<k}^m
H_{jk}^{(m)}(u,u_1,\ldots,u_m)
B_2(u)|\psi_{m-2}(u_1,\ldots,\hat{u}_j,\ldots,\hat{u}_k,\ldots,u_m)\rangle\,,
\ee
and
\be\label{doffshellTBV}
&&\langle\psi_m(u_1,\ldots,u_m)| t^{\left(1,1\right)}(u)=
\langle\psi_m(u_1,\ldots,u_m)|\Lambda(u,u_1,\ldots,u_m)\nonumber\\&&+
\sum_{j=1}^m \langle\psi_{m-1}(u_1,\ldots,\hat{u}_j,\ldots,u_m)|C_1(u)\prod_{p<j}^m\Omega(u_j,u_p)\frac{\bar E_j(u_j)Q_j(u_j-2\eta)}{Q_j(u_j)Q_j(u_j-\eta)}  F(u,u_j) \sinh(2\eta)\nonumber\\
&&+
\sum_{j=1}^m \langle\psi_{m-1}(u_1,\ldots,\hat{u}_j,\ldots,u_m)|C_3(u)\prod_{p<j}^m\Omega(u_j,u_p) \frac{\bar E_j(u_j)Q_j(u_j-2\eta)}{Q_j(u_j)Q_j(u_j-\eta)}\sinh(2\eta)\cosh(\eta)F(u,u_j+\eta)\nonumber\\
&&+
\sum_{j<k}^m
\langle\psi_{m-2}(u_1,\ldots,\hat{u}_j,\ldots,\hat{u}_k,\ldots,u_m)|C_2(u) H_{jk}^{(m)}(u,u_1,\ldots,u_m)\cosh^2(\eta)\,,
\ee
where
\begin{align}
\Lambda(u,u_1,\ldots,u_m) &=\Lambda_1(u)\frac{Q(u-2\eta)}{Q(u)}
   +\Lambda_2(u) 
   \frac{Q(u-2\eta)Q(u+\eta)}{Q(u)Q(u-\eta)}+\Lambda_3(u)\frac{Q(u+\eta)}{Q(u-\eta)}\,, \\
\bar E_j(u) &=
\Lambda_1(u)Q_j(u-\eta)-\Lambda_2(u)Q_j(u+\eta) \,,
\label{BPsingleT}
\end{align}
and
\be\label{defHjk}
&&H_{jk}^{(m)}(u,u_1,\ldots,u_m)\non\\
&&=4\cosh(\eta)\sinh^2(\eta)\frac{Q_j(u_j-2\eta)Q_{j,k}(u_k-2\eta)}{Q_j(u_j-\eta)Q_{j,k}(u_j)Q_{j,k}(u_k)Q_{j,k}(u_k-\eta)} \prod_{p<j}^m \Omega(u_j,u_p)  \prod_{q<k,q\neq j}^m \Omega(u_k,u_q)\non\\
&&\times
\left\{F(u,u_j) F(u,u_k) F(u_j,  u_k + \eta)
F(u_j, u_k - \eta) \bar E_j(u_j) \bar E_k(u_k) \right.\non\\&&\quad+\left. 
\sinh(\eta) F(u , u_k) F(u,  u_k + \eta) F(u_j , u_k) F(u_j,  u_k + \eta) Q_{j,k}(u_j+\eta)\Lambda_2(u_j)\bar E_k(u_k) \right.\non\\&&\quad+\left. 
\sinh(\eta) F(u , u_j) F(u,  u_j + \eta) F(u_j , u_k) F(u_j,  u_k - \eta) Q_{j,k}(u_k+\eta) \Lambda_2(u_k)\bar E_j(u_j)\right\}\,.
\ee
The set of polynomial equations
\be\label{periodicBET}
\bar E_j(u_j)=0\quad \textrm{for}\quad j=1,\dots,m\,,
\ee
are the Bethe ansatz equations from Tarasov's
construction. We remark that the unwanted terms have been written explicitly in
terms of the Bethe polynomials; in this way, it is clear that all unwanted terms disappear on-shell.

\subsection{Relating the two 
closed-chain Bethe 
ans\"atze}\label{sec:BArelations}

The Bethe equations from fusion 
(\ref{periodicBEfusion}) and from Tarasov's construction 
(\ref{periodicBET}) are equivalent. Indeed, since the factors in the 
Bethe polynomials
$E_j(u)$ (\ref{BPsingleF}) and $\bar E_j(u)$ (\ref{BPsingleT}) are related by
\be
\frac{\Lambda_{1}(u)}{\Lambda_{2}(u)} 
= \frac{\lambda_{1}(u+\tfrac{\eta}{2})}{\lambda_{2}(u+\tfrac{\eta}{2})} 
\,,
\ee
the solutions of the two sets of Bethe equations are related  
by $\tfrac{\eta}{2}$ shifts.

We claim that the relation between the off-shell fusion-Bethe vectors 
(\ref{FBVsingle}), (\ref{dFBVsingle}), and the off-shell Tarasov-Bethe vectors 
(\ref{TBVsingle}), (\ref{dTBVsingle}) is given by
\be\label{relationSR}
|\psi_m(u_1,\dots,u_m)\rangle =s(u_1,\dots,u_m)\Big|\phi_m\left(u_1+\frac{\eta}{2},\dots,u_m+\frac{\eta}{2}\right)\Big\rangle\,,
\ee
and
\be\label{drelationSR}
\langle\psi_m(u_1,\dots,u_m)| = \Big\langle\phi_m\left(u_1+\frac{\eta}{2},\dots,u_m+\frac{\eta}{2}\right)\Big| s(u_1,\dots,u_m)\cosh^m(\eta)\,,
\ee
where 
\be\label{defsfunc}
s(u_p,\dots,u_m)= 2^{\frac{m-(p-1)}{2}}\prod_{i=p}^m\lambda_1\left(u_i-\frac{\eta}{2}\right)\prod_{j,k=p,j<k}^m\frac{\sinh(u_j-u_k-2\eta)}{\sinh(u_j-u_k-\eta)}\,.
\ee
We prove (\ref{relationSR}) by induction; the proof of (\ref{drelationSR}) is similar. The cases $m=1$ and $m=2$ follow from brute force computation. We first note that
the representations (\ref{rep1}) and (\ref{rep2}) are connected by means
of the relation (\ref{SRmono22to12}), which gives us in particular (see 
e.g. \cite{Kitanine:2001bq})
\be\label{spin1to}
&&B_{1}(u)= \frac{1}{\sqrt{2}}\left(A\left(u+\frac{\eta }{2}\right)B\left(u-\frac{\eta}{2}\right)+B\left(u+\frac{\eta }{2}\right)A\left(u-\frac{\eta}{2}\right)\right)=\sqrt{2}B\left(u+\frac{\eta }{2}\right)A\left(u-\frac{\eta}{2}\right),
   \non\\
&&B_{2}(u)= B\left(u+\frac{\eta}{2}\right)B\left(u-\frac{\eta }{2}\right)\,,
\ee
and then use commutation relations from 
(\ref{RTT12}), the action (\ref{actionsinglerow}) and the identity
\be\label{lambdaidentity}
\Lambda_1(u)=\lambda_1\left(u+\frac{\eta}{2}\right)\lambda_1\left(u-\frac{\eta}{2}\right)\,.
\ee
Next, let us compute $|\psi_{m+1}(u_1,\dots,u_{m+1})\rangle$ supposing (\ref{relationSR}) is valid. By (\ref{TBVsingle}) we have
\be
&&|\psi_{m+1}(u_1,\dots,u_{m+1})\rangle=B_1(u_1)|\psi_{m}(u_2,\ldots,u_{m+1})\rangle\nonumber\\&&\quad\quad-
 B_2(u_1)\sum_{i=2}^{m+1} \gamma_i^{(m+1)}(u_1,\ldots,u_{m+1}) 
 |\psi_{m-1}(u_2,\ldots,\hat{u}_i,\ldots,u_{m+1})\rangle \,,
\ee
and, by means of (\ref{relationSR}), we obtain
\begin{align}
|\psi_{m+1}(u_1,\dots,u_{m+1})\rangle &=s(u_2,\ldots,u_{m+1})\, B_1(u_1)|\phi_{m}\left(\tilde{u}_2,\ldots,\tilde{u}_{m+1}\right)\rangle\nonumber\\
&\hspace{-1.5in} -
B_2(u_1)\sum_{i=2}^{m+1} \gamma_i^{(m+1)}(u_1,\ldots,u_{m+1}) 
s(u_2,\ldots,\hat{u}_i,\ldots,u_{m+1})|\phi_{m-1}(\tilde{u}_2,\ldots,\hat{\tilde{u}}_i,\ldots,\tilde{u}_{m+1})\rangle \,,\non\\
\end{align}
where we have introduced
\be\label{defsfuncnew1}
s(u_p,\dots,\hat{u}_l,\ldots,u_m)= 2^{\frac{m-p}{2}}\prod_{i=p,i\neq l}^m\lambda_1\left(u_i-\frac{\eta}{2}\right)
\prod_{j,k=p,j<k,j\neq l,k\neq l}^m\frac{\sinh(u_j-u_k-2\eta)}{\sinh(u_j-u_k-\eta)}
\ee
and the notation
\be
\tilde u = u +\frac{\eta}{2}\,.
\label{tildenotation}
\ee
Using (\ref{spin1to}) we express $B_1(u_1)$ and $B_2(u_1)$ in terms 
of $A(u_1)$ and $B(u_1)$, and compute
\be\label{B1phi}
B_1(u_1)|\phi_{m}\left(\tilde{u}_2,\ldots,\tilde{u}_{m+1}\right)\rangle&=&\sqrt{2}\,B\left(\tilde{u}_1\right)A\left(u_1-\frac{\eta}{2}\right)|\phi_{m}\left(\tilde{u}_2,\ldots,\tilde{u}_{m+1}\right)\rangle\,,
\ee
\begin{align}
    \label{B2phi}
B_2(u_1)|\phi_{m-1}(\tilde{u}_2,\ldots,\hat{\tilde{u}}_i,\ldots,\tilde{u}_{m+1})\rangle  &=
B\left(\tilde{u}_1\right)B\left(u_1-\frac{\eta }{2}\right)|\phi_{m-1}(\tilde{u}_2,\ldots,\hat{\tilde{u}}_i,\ldots,\tilde{u}_{m+1})\rangle
\non\\
& = B\left(u_1-\frac{\eta }{2}\right)|\phi_{m}(\tilde{u}_1,\tilde{u}_2,\ldots,\hat{\tilde{u}}_i,\ldots,\tilde{u}_{m+1})\rangle\,.
\end{align}
By means of
(\ref{RTT12}) and the action (\ref{actionsinglerow}), we obtain
\begin{align}
    \label{ABBB}
A(u)|\phi_{m}(u_1,\ldots,u_{m})\rangle&=
\lambda_1(u)\frac{Q(u-\eta)}{Q(u)}|\phi_{m}(u_1,\ldots,u_{m})\rangle\non\\
&\hspace{-1in}+
\sum_{j=1}^{m}\frac{\sinh(\eta)}{\sinh(u-u_j)}\lambda_1(u_j)\frac{Q_j(u_j-\eta)}{Q_j(u_j)}B(u)
|\phi_{m-1}(u_1,\ldots,\hat{u}_j,\ldots,u_{m})\rangle \,,
\end{align}
which leads to
\be\label{B1phiA}
&&B_1(u_1)\Big|\phi_{m}\left(\tilde{u}_2,\ldots,\tilde{u}_{m+1}\right)\Big\rangle=\sqrt{2}\,
\lambda_1\left(u_1-\frac{\eta}{2}\right)\prod_{j=2}^{m+1}\frac{\sinh(u_1-u_j-2\eta)}{\sinh(u_1-u_j-\eta)}|\phi_{m+1}(\tilde{u}_1,\tilde{u}_2,\ldots,\tilde{u}_{m+1})\rangle\non\\&&+
\sqrt{2}\,\sum_{j=2}^{m+1}\frac{\sinh(\eta)\lambda_1(\tilde{u}_j)}{\sinh(u_1-u_j-\eta)}\prod_{k=2,k\neq j}^{m+1}
\frac{\sinh(u_k-u_j+\eta)}{\sinh(u_k-u_j)}B\left(u_1-\frac{\eta}{2}\right)
|\phi_{m}(\tilde{u}_1,\tilde{u}_2,\ldots,\hat{\tilde{u}}_j,\ldots,\tilde{u}_{m+1})\rangle \,. \non\\
\ee
Using (\ref{B2phi}) and (\ref{B1phiA}) we obtain
\be
&&|\psi_{m+1}(u_1,\dots,u_{m+1})\rangle\non\\&&=s(u_2,\ldots,u_{m+1})
\sqrt{2}\,
\lambda_1\left(u_1-\frac{\eta}{2}\right)\prod_{j=2}^{m+1}\frac{\sinh(u_1-u_j-2\eta)}{\sinh(u_1-u_j-\eta)}
|\phi_{m+1}(\tilde{u}_1,\tilde{u}_2,\ldots,\tilde{u}_{m+1})\rangle\non\\
&&+
\sum_{j=2}^{m+1}
\Bigg\{s(u_2,\ldots,u_{m+1})\sqrt{2}\frac{\sinh(\eta)\lambda_1(\tilde{u}_j)}{\sinh(u_1-u_j-\eta)}\prod_{k=2,k\neq j}^{m+1}
\frac{\sinh(u_k-u_j+\eta)}{\sinh(u_k-u_j)}  \non\\&&\quad-\gamma_j^{(m+1)}(u_1,\ldots,u_{m+1}) 
s(u_2,\ldots,\hat{u}_j,\ldots,u_{m+1})\Bigg\}\non\\&&\quad\quad\times B\left(u_1-\frac{\eta}{2}\right)
|\phi_{m}(\tilde{u}_1,\tilde{u}_2,\ldots,\hat{\tilde{u}}_j,\ldots,\tilde{u}_{m+1})\rangle\,.
\ee
We can check that
\be
s(u_1,\ldots,u_{m+1}) = s(u_2,\ldots,u_{m+1})
\sqrt{2}\,
\lambda_1\left(u_1-\frac{\eta}{2}\right)\prod_{j=2}^{m+1}\frac{\sinh(u_1-u_j-2\eta)}{\sinh(u_1-u_j-\eta)} \,,
\ee
and, thanks to (\ref{lambdaidentity}),
\be
&&s(u_2,\ldots,u_{m+1})\sqrt{2}\frac{\sinh(\eta)\lambda_1(u_j+\frac{\eta}{2})}{\sinh(u_1-u_j-\eta)}\prod_{k=2,k\neq j}^{m+1}
\frac{\sinh(u_k-u_j+\eta)}{\sinh(u_k-u_j)}  \non\\
&&\quad=\gamma_j^{(m+1)}(u_1,\ldots,u_{m+1}) 
s(u_2,\ldots,\hat{u}_j,\ldots,u_{m+1}) \,,
\ee
which implies
\be
|\psi_{m+1}(u_1,\dots,u_{m+1})\rangle = s(u_1,\ldots,u_{m+1})\, 
|\phi_{m+1}(\tilde{u}_1,\tilde{u}_2,\ldots,\tilde{u}_{m+1})\rangle \,,
\ee 
and therefore completes the inductive proof of (\ref{relationSR}).
\qed

\subsection{Scalar products for Tarasov-Bethe vectors}\label{sec:SPT}

Combining the scalar product for the fusion-Bethe vectors 
(\ref{SLAVSR}) with the relations between Tarasov-Bethe vectors and 
fusion-Bethe vectors (\ref{relationSR}), (\ref{drelationSR}), 
it is now
straightforward to compute the scalar product between the off-shell state
$\langle\psi_m(u_1,\dots,u_m)|$ (\ref{dTBVsingle}) and the on-shell state
$|\psi_m(v_1,\dots,v_m)\rangle$ (\ref{TBVsingle}) 
\be\label{SLAVSRZF}
&&\langle\psi_m(u_1,\dots,u_m)|\psi_m(v_1,\dots,v_m)\rangle=\cosh^m(\eta)\, s(u_1,\dots,u_m)\, s(v_1,\dots,v_m)\non\\
&&\quad\quad\times\prod_{i=1}^m\lambda_2\left(u_i+\frac{\eta}{2}\right)
\frac{\textrm{det}_m\left(\frac{\partial}{\partial u_i}\lambda\left(v_j+\frac{\eta}{2},u_1+\frac{\eta}{2},\dots,u_m+\frac{\eta}{2}\right))\right)}
{\textrm{det}_m\left(F(v_i,u_j)\right)}\,,
\ee
which is in terms of the quantities $\lambda$ and $\lambda_{2}$ 
arising from fusion.
We can rewrite (\ref{SLAVSRZF}) in a similar way as in (\ref{SLAVBESR})
\be\label{SLAVBESRZF}
\langle\psi_m(u_1,\dots,u_m)|\psi_m(v_1,\dots,v_m)\rangle&=&\prod_{j<k}^m\frac{\sinh(u_j-u_k-2\eta)}{\sinh(u_j-u_k-\eta)}\frac{\sinh(v_j-v_k-2\eta)}{\sinh(v_j-v_k-\eta)}
\non\\&\times&
\prod_{i=1}^m
\Lambda_2(u_i)
\frac{\textrm{det}_m\left(\sinh(2\eta)F(u_i,v_j)F(v_j,u_i)\frac{\bar E_i(v_j)}{Q_i(v_j)}\right)}
{\textrm{det}_m\left(F(v_i,u_j)\right)}\,,\non\\
\ee
which is instead in terms of 
quantities arising from Tarasov's construction, namely,
the Bethe ansatz polynomial (\ref{periodicBET}) and the Baxter polynomial (\ref{BaxterQ}). The Slavnov formula
written in (\ref{SLAVBESR}) or 
(\ref{SLAVBESRZF}) therefore seems more ``universal'' than the formula given in terms of the eigenvalue
of the transfer matrix.


\section{Bethe ansatz: double-row transfer matrix}\label{sec:double}

In this section we obtain the eigenvectors of the double-row transfer matrices (\ref{DRtransfer12}) and (\ref{DRtransfer22}).

\subsection{Bethe vectors from fusion}

We firstly apply the Bethe ansatz to (\ref{DRtransfer12}), see \cite{Sklyanin:1988yz,Mezincescu:1990fc}. For that, we use the 2-dimensional auxiliary
space representation, \textit{i.e.}, we set
\be\label{rep1DR}
U_{a}^{(\tfrac{1}{2},1)}(u) = \left(
\begin{array}{cc}
 {\mA}(u) & {\mB}(u) \\
 {\mC}(u) & {\mD}(u)+\frac{\sinh(\eta)}{\sinh(2u+\eta)} \mA(u)
\end{array}
\right)_a
\ee
in which each entry is an operator acting on the quantum space ${\left(\mathbb{C}^{3}\right)}^{\otimes N}$. These operators
satisfy exchange relations dictated by (\ref{BYBEU1}).
The action of the double-row monodromy operators on the 
reference states (\ref{refstate},\ref{drefstate}) is given by
\be
\mA(u)\refs = \delta_1(u) \refs\,,\quad \mD(u)\refs = \delta_2(u) \refs\,,\quad \mC(u)\refs=0\,,
\ee
\be
\drefs \mA(u) =\drefs \delta_1(u) \,,\quad \drefs \mD(u) =\drefs \delta_2(u) \,,\quad \drefs \mB(u)=0\,,
\ee
where
\be\label{delta12}
&&\delta_1(u)=\sinh(u+\xi^{-})\sinh^{2N}\left(u+\frac{3\eta}{2}\right)\,,\non\\
&&\delta_2(u)= \frac{\sinh(2u)\sinh(\xi^{-}-u-\eta)}{\sinh(2u+\eta)}\sinh^{2N}\left(u-\frac{\eta}{2}\right)\,.
\ee
The transfer matrix (\ref{DRtransfer12}) is given by
\be
\tau^{(\tfrac{1}{2},1)}(u)=\frac{\sinh (2 (u+\eta )) \sinh 
(u-\xi^{+}) }{\sinh (2 u+\eta )}\mA(u)-\sinh (u+\eta+\xi^{+})\, \mD(u)\,.
\ee
The fusion-Bethe vectors are given by
\be
|\Phi_m(u_1,\dots,u_m)\rangle =\mB(u_1)\dots \mB(u_m)\refs\,,
\label{FBVdouble}
\ee
and
\be
\langle\Phi_m(u_1,\dots,u_m)|=\drefs \mC(u_1)\dots \mC(u_m)\,.
\label{dFBVdouble}
\ee
In fact, 
using the reflection algebra (\ref{BYBEU1})
we can show that
\be
&&\tau^{\left(\frac{1}{2},1\right)}(u)|\Phi_m(u_1,\dots,u_m)\rangle=
\delta(u,u_1,\dots, u_m)|\Phi_m(u_1,\dots,u_m)\rangle\nonumber\\&&\quad+
\sum_{j=1}^m \sinh(\eta) \sinh(2 
(u+\eta))\frac{\mathcal{F}\left(u,u_j\right)}{\sinh(2u_j+\eta)}\frac{\mE_j(u_j)}{\mQ_j(u_j)}\mB(u)|\Phi_{m-1}(u_1,\ldots,\hat{u}_j,\ldots,u_m)\rangle\,, \non\\
\ee
and
\be
&&\langle\Phi_m(u_1,\dots,u_m)| \tau^{\left(\frac{1}{2},1\right)}(u) =
\langle\Phi_m(u_1,\dots,u_m)| \delta(u,u_1,\dots u_m)\nonumber\\&&\quad+
\sum_{j=1}^m \langle\Phi_m (u_1,\ldots,\hat{u}_j,\ldots,u_m)|\mC(u) 
\frac{\mE_j(u_j)}{\mQ_j(u_j)}\frac{\mathcal{F}\left(u,u_j\right)}{\sinh(2u_j+\eta)}\sinh(2 (u+\eta))\sinh(\eta)\,, \non\\
\ee
where
\begin{align}
\delta(u,u_1,\dots, u_m)&=\frac{\sinh(2(u+\eta))\sinh(u-\xi^{+})}{\sinh(2u+\eta)}\delta_1(u)\frac{\mQ(u-\eta)}{\mQ(u)}\non\\
&\quad -\sinh(u+\eta+\xi^{+})\delta_2(u)\frac{\mQ(u+\eta)}{\mQ(u)}\,, \\
\mE_j(u)&=\sinh(2u)\sinh(u-\xi^{+})\delta_1(u)\mQ_j(u-\eta)\non\\
&\quad +\sinh(2u+\eta)\sinh(u+\eta+\xi^{+})\delta_2(u)\mQ_j(u+\eta)
\label{BPdoubleF}
\end{align}
and
\be
\mF(u,v)=\frac{1}{\sinh(u-v)\sinh(u+v+\eta)}\,.
\ee
In the above formulae we have introduced the double-row Baxter Q-polynomial,
\be\label{BaxterQDR}
\mQ(u)=\prod_{i=1}^m \sinh(u-u_i)\sinh(u+u_i+\eta)\,,
\ee
and the indexed double-row Baxter Q-polynomial,
\be
\mQ_j(u)=\prod_{i\neq j}^m \sinh(u-u_i)\sinh(u+u_i+\eta)\,.
\ee
The set of polynomial equations
\be\label{openBEfusion}
\mE_j(u_j)=0\quad \textrm{for}\quad j=1,\dots,m\,,
\ee
are known as the Bethe ansatz equations for the double-row transfer matrix. Again, we will refer to $\mE_j(u)$ as the Bethe ansatz polynomials for the double-row transfer matrix.

Finally, let us suppose that $\{u_1,\dots,u_m\}$ satisfy the Bethe equations (\ref{openBEfusion}) (\textit{i.e.}, they are on-shell rapidities) and that there is no restriction on 
$\{v_1,\dots,v_m\}$ (\textit{i.e.}, they are off-shell rapidities). Then, the scalar product $\langle\Phi_m(u_1,\dots,u_m)|\Phi_m(v_1,\dots,v_m)\rangle$ is given by the compact formula \cite{Kitanine:2007bi},
\be\label{SLAVDR}
\langle\Phi_m(u_1,\dots,u_m)|\Phi_m(v_1,\dots,v_m)\rangle&=&
\prod_{i=1}^m\frac{\delta_2(u_i)}{\sinh(2(v_i+\eta))\sinh(u_i-\xi^{+})}\prod_{j<i}^m\frac{\sinh(u_i+u_j+2\eta)}{\sinh(u_i+u_j)}
\non\\&\times&\frac{\textrm{det}_m\left(\frac{\partial}{\partial u_i}\delta(v_j,u_1,\dots,u_m)\right)}
{\textrm{det}_m\left(\mF(v_i,u_j)\right)}\,,
\ee
which is the Slavnov formula for the double-row transfer matrix. Using
\be
\frac{\partial}{\partial u_i}\delta(v_j,u_1,\dots,u_m)=\sinh(\eta)\sinh(2u_i+\eta)\frac{\sinh(2(v_j+\eta))}{\sinh(2v_j+\eta)}\mF(u_i,v_j)\mF(v_j,u_i)\frac{\mE_i(v_j)}{\mQ_i(v_j)}
\ee
we can rewrite the Slavnov formula as
\be\label{SLAVBEDR}
\langle\Phi_m(u_1,\dots,u_m)|\Phi_m(v_1,\dots,v_m)\rangle&=&
\prod_{i=1}^m\frac{\sinh(2u_i+\eta)\delta_2(u_i)}{\sinh(2v_i+\eta)\sinh(u_i-\xi^{+})}\prod_{j<i}^m\frac{\sinh(u_i+u_j+2\eta)}{\sinh(u_i+u_j)}
\non\\&\times&
\frac{\textrm{det}_m\left(\sinh(\eta)\mF(u_i,v_j)\mF(v_j,u_i)\frac{\mE_i(v_j)}{\mQ_i(v_j)}\right)}
{\textrm{det}_m\left(\mF(v_i,u_j)\right)}\,,\non\\
\ee
which is given in terms of the double-row Bethe ansatz and Baxter polynomials, similarly to
the single-row case.

\subsection{Bethe vectors from Tarasov's construction}

We now apply the Bethe ansatz to (\ref{DRtransfer22}), see \cite{Fan:1997wpk,Kurak:2004ip}. Here, we use the 3-dimensional auxiliary
space representation, \textit{i.e.}, we set
\be\label{rep2DR}
&&U^{\left(1,1\right)}_a(u)=\non\\
&&\left(
\begin{array}{ccc}
 \mA_1(u) & \mB_1(u) & \mB_2(u) \\
 \mC_1(u) & \mA_2(u)+\frac{\sinh (2 \eta )}{\sinh (2 (u+\eta))}\mA_1(u) & \mB_3(u) \\
 \mC_2(u) & \mC_3(u) & \mA_3(u)+\frac{\sinh (\eta ) \sinh (2 \eta)}{\sinh (2 (u+\eta )) \sinh (2 u+\eta )}\mA_1(u)+\frac{\sinh (2\eta )}{\sinh (2u)}\mA_2(u) \\
\end{array}
\right)_a\,,\non\\
\ee
where each entry acts on the quantum space
${\left(\mathbb{C}^{3}\right)}^{\otimes N}$.  These operators satisfy
exchange relations dictated by (\ref{BYBEU2}).  The action of the
monodromy operators (\ref{rep2DR}) on the reference state
(\ref{refstate},\ref{drefstate}) is given by
\be\label{actionref1DR}
\mA_j(u)\refs = \Delta_j(u)\refs\,,\quad \mC_j(u)\refs = 0\,,
\ee
\be\label{dactionref1DR}
\drefs \mA_j(u)=\drefs \Delta_j(u)\,,\quad \drefs \mB_j(u)=0\,,
\ee
for $j=1,2,3$ and where
\begin{align}
\Delta_1(u) &=\cosh \left(u+\frac{\eta }{2}\right) \sinh \left(u-\frac{\eta }{2}+\xi ^-\right) \sinh
   \left(u+\frac{\eta }{2}+\xi ^-\right)\left(\sinh (u+\eta) \sinh (u+2 \eta)\right)^{2N}\,,\non\\
   \Delta_2(u) &=-\frac{\cosh \left(u+\frac{\eta }{2}\right) \sinh (2 u) \sinh \left(u+\frac{3 \eta }{2}-\xi ^-\right)
   \sinh \left(u-\frac{\eta }{2}+\xi ^-\right)}{\sinh (2 (u+\eta ))}\left(\sinh (u) \sinh (u+\eta )\right)^{2N}\,,\non\\
\Delta_3(u) &=\frac{\sinh (2 u-\eta ) \sinh \left(u+\frac{\eta }{2}-\xi ^-\right) \sinh \left(u+\frac{3 \eta }{2}-\xi
   ^-\right)}{2 \sinh \left(u+\frac{\eta }{2}\right)}\left(\sinh (u) \sinh (u-\eta )\right)^{2N}\,.\non
\label{Delta123}\\
\end{align}
The transfer matrix (\ref{DRtransfer22}) is given by
\begin{align}
\tau^{(1,1)}(u)&=-\frac{\sinh (2 u+3 \eta ) \sinh \left(u-\frac{\eta }{2}-\xi^{+}\right) \sinh \left(u+\frac{\eta}{2}-\xi^{+}\right) }{2 \sinh \left(u+\frac{\eta
   }{2}\right)}\mA_1(u)\non\\
   &+  \frac{\cosh \left(u+\frac{\eta }{2}\right)  \sinh (2(u+\eta) ) \sinh\left(u-\frac{\eta }{2}-\xi^{+}\right) \sinh \left(u+\frac{3 \eta }{2}+\xi^{+}\right)
   }{\sinh (2u)}\mA_2(u)\non\\
   &-\cosh \left(u+\frac{\eta }{2}\right) \sinh\left(u+\frac{\eta }{2}+\xi^{+}\right) \sinh \left(u+\frac{3 \eta }{2}+\xi^{+}\right)
   \mA_3(u)\,.
\end{align}
The Bethe vector is constructed by means of the recursion relation
\begin{align}\label{TBVdouble}
|\Psi_m(u_1,\dots,u_m)\rangle&=\mB_1(u_1)|\Psi_{m-1}(u_2,\ldots,u_m)\rangle\nonumber\\
&-\mB_2(u_1)\sum_{i=2}^m \Gamma_i^{(m)}(u_1,\ldots,u_m) 
|\Psi_{m-2}(u_2,\ldots,\hat{u}_i,\ldots,u_m)\rangle \,,
\end{align}
where
\be\label{defGamma}
&&\Gamma_i^{(m)}(u_1,\ldots,u_m)=2\sinh(\eta)\frac{\bar\mQ_{1,i}(u_i-2\eta)}{\bar\mQ_{1,i}(u_i)}\prod_{j=2,j<i}^m \Omega(u_i,u_j)
\non\\&&\quad\times\left(\frac{\sinh(2u_i)}{\sinh(u_1-u_i-\eta)\sinh(2(u_i+\eta))}\Delta_1(u_i)-\frac{1}{\sinh(u_1+u_i+\eta)}\frac{\bar\mQ_{1,i}(u_i+\eta)}{\bar\mQ_{1,i}(u_i-\eta)}\Delta_2(u_i) \right)\,,\non\\
\ee
and where we have introduced a new Baxter double-row Q-polynomial
\be\label{BaxterQbar}
\bar\mQ(u)=\prod_{i=1}^m{\sinh(u-u_i)\sinh(u+u_i+2\eta)}\,,
\ee
\be\label{BaxterQbarj}
\bar\mQ_j(u)=\prod_{i\neq j}^m{\sinh(u-u_i)\sinh(u+u_i+2\eta)}\,,
\ee
\be\label{BaxterQbarjk}
\bar\mQ_{j,k}(u)=\prod_{i\neq j,k}^m{\sinh(u-u_i)\sinh(u+u_i+2\eta)}\,.
\ee
The dual Bethe vector is given by
\begin{align}\label{dTBVdouble}
\langle\Psi_m(u_1,\dots,u_m)|&=\langle\Psi_{m-1}(u_2,\ldots,u_m)| \mC_1(u_1)\nonumber\\
&-\sum_{i=2}^m \tilde\Gamma_i^{(m)}(u_1,\ldots,u_m) 
\langle\Psi_{m-2}(u_2,\ldots,\hat{u}_i,\ldots,u_m)|  \mC_2(u_1) \,,	
\end{align}
where
\be
&&\tilde\Gamma_i^{(m)}(u_1,\ldots,u_m)=\sinh(2\eta)\cosh(\eta)\frac{\bar\mQ_{1,i}(u_i-2\eta)}{\bar\mQ_{1,i}(u_i)}\prod_{j=2,j<i}^m \Omega(u_i,u_j)\non\\
&&\quad\times\left(\frac{\sinh(2u_i)}{\sinh(u_1-u_i-\eta)\sinh(2(u_i+\eta))}\Delta_1(u_i)-\frac{1}{\sinh(u_1+u_i+\eta)}\frac{\bar\mQ_{1,i}(u_i+\eta)}{\bar\mQ_{1,i}(u_i-\eta)}\Delta_2(u_i) \right)\,.\non\\
\ee
Notice that the the initial conditions
\be
|\Psi_{0}\rangle = \refs\,,\quad \langle\Psi_{0}|=\drefs
\ee
are assumed for (\ref{TBVdouble}) and (\ref{dTBVdouble}).

Using the reflection algebra (\ref{BYBEU2}) one can show that the vectors (\ref{TBVdouble}) and (\ref{dTBVdouble}) satisfy the off-shell equations
\be\label{offshellTBVDR}
&& \tau^{\left(1,1\right)}(u)|\Psi_m(u_1,\ldots,u_m)\rangle =
\Delta(u,u_1,\ldots,u_m)|\Psi_m(u_1,\ldots,u_m)\rangle\nonumber\\&&+
\sum_{j=1}^m \mathcal{G}_1(u,u_j) \frac{\bar \mE_j(u_j)\bar\mQ_j(u_j-2\eta)}{\bar\mQ_j(u_j)\bar\mQ_j(u_j-\eta)}\prod_{p<j}^m\Omega(u_j,u_p)\mB_1(u)|\Psi_{m-1}(u_1,\ldots,\hat{u}_j,\ldots,u_m)\rangle\nonumber\\
&&+
\sum_{j=1}^m \mathcal{G}_2(u,u_j) \frac{\bar \mE_j(u_j)\bar\mQ_j(u_j-2\eta)}{\bar\mQ_j(u_j)\bar\mQ_j(u_j-\eta)}\prod_{p<j}^m\Omega(u_j,u_p)\mB_3(u)|\Psi_{m-1}(u_1,\ldots,\hat{u}_j,\ldots,u_m)\rangle\nonumber\\
&&+
\sum_{j<k}^m
\mH_{jk}^{(m)}(u,u_1,\ldots,u_m)
\mB_2(u)|\Psi_{m-2}(u_1,\ldots,\hat{u}_j,\ldots,\hat{u}_k,\ldots,u_m)\rangle\,,
\ee
and
\be\label{doffshellTBVDR}
&&\langle\Psi_m(u_1,\ldots,u_m)| \tau^{\left(1,1\right)}(u)=
\langle\Psi_m(u_1,\ldots,u_m)|\Delta(u,u_1,\ldots,u_m)\nonumber\\&&+
\sum_{j=1}^m \langle\Psi_{m-1}(u_1,\ldots,\hat{u}_j,\ldots,u_m)|\mC_1(u)\prod_{p<j}^m\Omega(u_j,u_p)\frac{\bar \mE_j(u_j)\bar\mQ_j(u_j-2\eta)}{\bar\mQ_j(u_j)\bar\mQ_j(u_j-\eta)}  \mathcal{G}_1(u,u_j)\nonumber\\
&&+
\sum_{j=1}^m \langle\Psi_{m-1}(u_1,\ldots,\hat{u}_j,\ldots,u_m)|\mC_3(u)\prod_{p<j}^m\Omega(u_j,u_p) \frac{\bar \mE_j(u_j)\bar\mQ_j(u_j-2\eta)}{\bar\mQ_j(u_j)\bar\mQ_j(u_j-\eta)}\mathcal{G}_2(u,u_j)\cosh^2(\eta)\nonumber\\
&&+
\sum_{j<k}^m
\langle\Psi_{m-2}(u_1,\ldots,\hat{u}_j,\ldots,\hat{u}_k,\ldots,u_m)|\mC_2(u) \mH_{jk}^{(m)}(u,u_1,\ldots,u_m)\cosh^2(\eta)\,,
\ee
where
\be
&&\Delta(u,u_1,\ldots,u_m)=-\frac{\sinh (2 u+3 \eta ) \sinh \left(u-\frac{\eta }{2}-\xi^{+}\right) \sinh \left(u+\frac{\eta}{2}-\xi^{+}\right) }{2 \sinh \left(u+\frac{\eta
   }{2}\right)}\Delta_1(u)\frac{\bar\mQ(u-2\eta)}{\bar\mQ(u)}\non\\&&+
   \frac{\cosh \left(u+\frac{\eta }{2}\right)  \sinh (2(u+\eta) ) \sinh\left(u-\frac{\eta }{2}-\xi^{+}\right) \sinh \left(u+\frac{3 \eta }{2}+\xi^{+}\right)
   }{\sinh (2u)}\Delta_2(u) \frac{\bar\mQ(u-2\eta)\bar\mQ(u+\eta)}{\bar\mQ(u)\bar\mQ(u-\eta)}
\non\\&&-\cosh \left(u+\frac{\eta }{2}\right) \sinh\left(u+\frac{\eta }{2}+\xi^{+}\right) \sinh \left(u+\frac{3 \eta }{2}+\xi^{+}\right)
   \Delta_3(u)\frac{\bar\mQ(u+\eta)}{\bar\mQ(u-\eta)}\,,
\ee
\begin{align}
\label{BPdoubleT}
\bar \mE_j(u)&=
\sinh(2u)\sinh\left(u+\frac{\eta}{2}-\xi^+\right)\Delta_1(u)\bar\mQ_j(u-\eta)\non\\
&+\sinh(2(u+\eta))\sinh\left(u+\frac{3\eta}{2}+\xi^+\right)\Delta_2(u)\bar\mQ_j(u+\eta)
\,,	
\end{align}
\be\label{defHjk}
&&\mH_{jk}^{(m)}(u,u_1,\ldots,u_m)\non\\
&&=\frac{\bar\mQ_j(u_j-2\eta)\bar\mQ_{j,k}(u_k-2\eta)}{\bar\mQ_j(u_j-\eta)\bar\mQ_{j,k}(u_j)\bar\mQ_{j,k}(u_k)\bar\mQ_{j,k}(u_k-\eta)} \prod_{p<j}^m \Omega(u_j,u_p)  \prod_{q<k,q\neq j}^m \Omega(u_k,u_q)\non\\
&&\times
\left\{\beta_1(u,u_j,u_k) \bar E_j(u_j) \bar E_k(u_k) + 
\beta_2(u,u_k,u_j) \bar\mQ_{j,k}(u_j+\eta)\Lambda_2(u_j)\bar E_k(u_k) \right.\non\\&&\left. 
+\beta_2(u,u_j,u_k)\bar\mQ_{j,k}(u_k+\eta) \Lambda_2(u_k)\bar E_j(u_j)\right\}\,,
\ee
and the functions $\mathcal{G}_1(u,u_j)$, $\mathcal{G}_2(u,u_j)$, $\beta_1(u,u_j,u_k)$ and $\beta_2(u,u_j,u_k)$ are given in the appendix \ref{sec:auxiliary}. 

The set of polynomial equations
\be\label{doubleBET}
\bar \mE_j(u_j)=0\quad \textrm{for}\quad j=1,\dots,m\,,
\ee
are the Bethe ansatz equations for the double-row transfer matrix 
from Tarasov's construction. We remark that the unwanted terms have been written explicitly in
terms of the Bethe polynomials; in this way, it is clear that all unwanted terms disappear on-shell.

\subsection{Relating 
the two open-chain Bethe ans\"atze}\label{sec:BArelationsDR}

As in the periodic case, the open-chain Bethe equations from fusion 
(\ref{openBEfusion}) and from Tarasov's construction 
(\ref{doubleBET}) are equivalent. Indeed, since the factors in the 
Bethe polynomials
$\mE_j(u)$ (\ref{BPdoubleF}) and $\bar \mE_j(u)$ (\ref{BPdoubleT}) are related by
\be
\frac{\sinh(2u)\sinh\left(u+\frac{\eta}{2}-\xi^+\right)\Delta_1(u)}
{\sinh(2(u+\eta))\sinh\left(u+\frac{3\eta}{2}+\xi^+\right)\Delta_2(u)}=
\left(\frac{\sinh(2u)\sinh(u-\xi^{+})\delta_1(u)}{\sinh(2u+\eta)\sinh(u+\eta+\xi^{+})\delta_2(u)}
\right)\Big\vert_{u\mapsto u + \frac{\eta}{2}}\,, \non \\
\ee
the solutions of the two sets of Bethe equations are related  
by $\tfrac{\eta}{2}$ shifts .

We claim that the off-shell Tarasov-Bethe vectors (\ref{TBVdouble}), 
(\ref{dTBVdouble}) are related to the off-shell 
fusion-Bethe vectors (\ref{FBVdouble}), (\ref{dFBVdouble}) by
\be\label{relationDR}
|\Psi_m(u_1,\dots,u_m)\rangle =\bar s(u_1,\dots,u_m)\Big|\Phi_m\left(u_1+\frac{\eta}{2},\dots,u_m+\frac{\eta}{2}\right)\Big\rangle\,,
\ee
and
\be\label{drelationDR}
\langle\Psi_m(u_1,\dots,u_m)| = \Big\langle\Phi_m\left(u_1+\frac{\eta}{2},\dots,u_m+\frac{\eta}{2}\right)\Big| \bar s(u_1,\dots,u_m)\cosh^m(\eta)\,,
\ee
where
\be\label{defsfuncDR}
\bar s(u_p,\dots,u_m)= 
2^{-\frac{m-p+1}{2}}\prod_{i=p}^m\frac{\sinh(2u_i)\delta_1\left(u_i-\frac{\eta}{2}\right)}{\sinh\left(u_i+\frac{\eta}{2}\right)}
\prod_{j=k=p, j<k}^m\frac{\sinh(u_j+u_k)\sinh(u_j-u_k-2\eta)}{\sinh(u_j+u_k+\eta)\sinh(u_j-u_k-\eta)}\,. \non\\
\ee
We now prove (\ref{relationDR}) by induction; the proof of 
(\ref{drelationDR}) is similar. The cases $m=1$ and $m=2$ can be 
proved by brute-force computation.  We begin by using 
(\ref{TBVdouble}) to obtain
\begin{align}
|\Psi_{m+1}(u_1,\dots,u_{m+1})\rangle&=\mB_1(u_1)|\Psi_{m}(u_2,\ldots,u_{m+1})\rangle\nonumber\\
&\hspace{-0.5in}-\mB_2(u_1)\sum_{i=2}^{m+1} \Gamma_i^{(m+1)}(u_1,\ldots,u_{m+1}) 
|\Psi_{m-1}(u_2,\ldots,\hat{u}_i,\ldots,u_{m+1})\rangle \,.
\end{align}
The induction hypothesis (\ref{relationDR}) implies
\begin{align}
|\Psi_{m+1}(u_1,\dots,u_{m+1})\rangle &= \bar{s}(u_2,\dots,u_{m+1})\, \mB_1(u_1)\, |\Phi_m\left(\tilde{u}_2,\dots,\tilde{u}_{m+1}\right)\rangle
\\
&\hspace{-1.5in}-\mB_2(u_1)\sum_{i=2}^{m+1} \Gamma_i^{(m+1)}(u_1,\ldots,u_{m+1})\,  
\bar{s}(u_2,\dots, \hat{u}_{i}, \ldots, u_{m+1})\,
|\Phi_{m-1}(\tilde{u}_2,\ldots,\hat{\tilde{u}}_i,\dots,\tilde{u}_{m+1}) \rangle \,, \non
\end{align}
where we have introduced
\be
\lefteqn{\bar s(u_p,\dots,\hat{u}_{l}, \ldots,u_m) }\\ 
&&=2^{-\frac{m-p}{2}}\prod_{i=p, i\ne 
l}^m\frac{\sinh(2u_i)\delta_1\left(u_i-\frac{\eta}{2}\right)}{\sinh\left(u_i+\frac{\eta}{2}\right)} 
\prod_{j=k=p, j<k, j\ne l, k\ne 
l}^m\frac{\sinh(u_j+u_k)\sinh(u_j-u_k-2\eta)}{\sinh(u_j+u_k+\eta)\sinh(u_j-u_k-\eta)}\,,  \non
\ee
and we have again employed the notation (\ref{tildenotation}). 
We now use the relations
\begin{align}\label{spin1toDR}
\mB_{1}(u) &= \frac{\sqrt{2}\cosh \left(u+\frac{\eta }{2}\right) }{2}\mathcal{A}\left(u+\frac{\eta}{2}\right)\mathcal{B}\left(u-\frac{\eta}{2}\right)+
\frac{\sqrt{2}\sinh\left(\eta\right)}{4 \sinh \left(u+\frac{\eta}{2}\right)}
\mathcal{B}\left(u+\frac{\eta}{2}\right)\mathcal{D}\left(u-\frac{\eta }{2}\right)\non\\
& + \frac{\sqrt{2} \left(\sinh ^2(2u)+\sinh ^2(\eta )\right)}{4 \sinh (2u) \sinh\left(u+\frac{\eta }{2}\right)}\mathcal{B}\left(u+\frac{\eta}{2}\right)\mathcal{A}\left(u-\frac{\eta }{2}\right)
   \non\\
 &= \frac{\sinh (2u)}{\sqrt{2}\sinh\left(u+\frac{\eta}{2}\right)}\mathcal{B}\left(u+\frac{\eta}{2}\right)\mathcal{A}\left(u-\frac{\eta }{2}\right)  \,, \\   
\mB_{2}(u) &= \frac{\sinh (2u)}{2\sinh\left(u+\frac{\eta }{2}\right)}\mathcal{B}\left(u+\frac{\eta}{2}\right)\mathcal{B}\left(u-\frac{\eta }{2}\right) 
  \,, 
\end{align}
which follow from (\ref{Ufusion}) and the reflection algebra 
(\ref{BYBEU1}). In this way, we obtain
\begin{align}
|\Psi_{m+1}(u_1,\dots,u_{m+1})\rangle &= \frac{\sinh 
(2u_{1})}{2\sinh\left(u_{1}+\frac{\eta}{2}\right)}\Bigg\{
\sqrt{2}\bar{s}(u_2,\dots,u_{m+1})\, \mathcal{B}(\tilde{u}_1)\, 
\mathcal{A}(u_1-\frac{\eta}{2})\, |\Phi_m\left(\tilde{u}_2,\dots,\tilde{u}_{m+1}\right)\rangle
\nonumber\\
&\hspace{-1.5in}-\sum_{i=2}^{m+1} \Gamma_i^{(m+1)}(u_1,\ldots,u_{m+1})\,  
\bar{s}(u_2,\dots, \hat{u}_{i}, \ldots, u_{m+1})\,
\mathcal{B}(\tilde{u}_1)\, \mathcal{B}(u_1-\frac{\eta}{2})\, 
|\Phi_{m-1}(\tilde{u}_2,\ldots,\hat{\tilde{u}}_i,\dots,\tilde{u}_{m+1}) \rangle \Bigg\}\,.
\end{align}
Using the result
\begin{align}
\mathcal{A}(u)\, |\Phi_{m}(u_1,\dots,u_m)\rangle  & = \delta_{1}(u) 
\frac{\mQ(u-\eta)}{\mQ(u)}  |\Phi_{m}(u_1,\dots,u_m)\rangle \non\\
&\hspace{-1.5in}+\sum_{j=1}^{m} \Big[ \frac{\sinh(\eta)\, \sinh(2 
u_{j})}{\sinh(u-u_{j})\, \sinh(2u_{j}+\eta)}  \delta_{1}(u_{j}) 
\frac{\mQ_{j}(u_{j}-\eta)}{\mQ_{j}(u_{j})} \non \\
&\hspace{-1.5in}-\frac{\sinh(\eta)}{\sinh(u+u_{j}+\eta)}  \delta_{2}(u_{j}) 
\frac{\mQ_{j}(u_{j}+\eta)}{\mQ_{j}(u_{j})} \Big]\mathcal{B}(u)\, 
|\Phi_{m-1}(u_1,\dots,\hat{u}_{j}, \ldots, u_m)\rangle 
\end{align}
that can be obtained using the reflection algebra (\ref{BYBEU1}), we 
find that
\begin{align}
|\Psi_{m+1}(u_1,\dots,u_{m+1})\rangle &= \frac{\sinh 
(2u_{1})}{2\sinh\left(u_{1}+\frac{\eta}{2}\right)}
\sqrt{2}\bar{s}(u_2,\dots,u_{m+1})\, 
\delta_{1}(u_{1}-\frac{\eta}{2})\, \non\\
& \times \prod_{k=2}^{m+1}\frac{\sinh(u_{1}-u_{k}-2\eta)\, 
\sinh(u_{1}+u_{k})}{\sinh(u_{1}-u_{k}-\eta)\, \sinh(u_{1}+u_{k}+\eta)}
\, |\Phi_{m+1}\left(\tilde{u}_1,\dots,\tilde{u}_{m+1}\right)\rangle \non\\
& +\frac{\sinh(2u_{1})}{2\sinh\left(u_{1}+\frac{\eta}{2}\right)}
\sum_{j=2}^{m+1}\Bigg\{  \alpha_{j}(u_{1}, \ldots, u_{m+1})
\non\\
&\hspace{-1.5in}
- \Gamma_j^{(m+1)}(u_1,\ldots,u_{m+1})\,  
\bar{s}(u_2,\dots, \hat{u}_{j}, \ldots, u_{m+1}) \Bigg\}
\mathcal{B}(u_1-\frac{\eta}{2})\, 
|\Phi_{m}(\tilde{u}_1,\ldots,\hat{\tilde{u}}_j,\dots,\tilde{u}_{m+1}) 
\rangle \,,
\end{align}
where
\begin{align}
    \alpha_{j}(u_{1}, \ldots, u_{m+1}) &= \sqrt{2}\bar{s}(u_2,\dots,u_{m+1})\, 
\sinh(\eta) \non\\
&\hspace{-1.5in} \times \Bigg[
\frac{\sinh(2\tilde{u}_{j})\, \delta_{1}(\tilde{u}_{j})}{\sinh(u_{1}-\tilde{u}_{j}-\frac{\eta}{2})\, \sinh(2\tilde{u}_{j}+\eta)} 
\prod_{k=2,k\neq j}^{m+1}
\frac{\sinh(u_j-u_k-\eta)\sinh(u_j+u_k+\eta)}{\sinh (u_j-u_k)\sinh 
(u_j+u_k+2 \eta)} \non\\
&\hspace{-1.5in} -\frac{\delta_{2}(\tilde{u}_{j})}{\sinh(u_{1}+\tilde{u}_{j}+\frac{\eta}{2})} 
\prod_{k=2,k\neq j}^{m+1}
\frac{\sinh(u_j-u_k+\eta)\sinh(u_j+u_k+3\eta)}{\sinh(u_j-u_k)\sinh(u_j+u_k+2\eta)} \Bigg] \,.
\end{align}
Finally, using the identities
\begin{align}
\bar{s}(u_1,\dots,u_{m+1}) &=  \frac{\sinh 
(2u_{1})}{2\sinh\left(u_{1}+\frac{\eta}{2}\right)}
\sqrt{2}\bar{s}(u_2,\dots,u_{m+1})\, 
\delta_{1}(u_{1}-\frac{\eta}{2})\, \non\\
& \times \prod_{k=2}^{m+1}\frac{\sinh(u_{1}-u_{k}-2\eta)\, 
\sinh(u_{1}+u_{k})}{\sinh(u_{1}-u_{k}-\eta)\, \sinh(u_{1}+u_{k}+\eta)}
\end{align}
and
\be
\alpha_{j}(u_{1}, \ldots, u_{m+1}) = \Gamma_j^{(m+1)}(u_1,\ldots,u_{m+1})\,  
\bar{s}(u_2,\dots, \hat{u}_{j}, \ldots, u_{m+1}) \,, \non
\ee
we conclude that
\be
|\Psi_{m+1}(u_1,\dots,u_{m+1})\rangle = \bar{s}(u_1,\ldots,u_{m+1})\, 
|\Phi_{m+1}(\tilde{u}_1,\ldots,\tilde{u}_{m+1})\rangle \,,
\ee 
which completes the inductive proof of (\ref{relationDR}).
\qed

\subsection{Scalar products for Tarasov-Bethe vectors}

Combining the scalar product for the fusion-Bethe vectors 
(\ref{SLAVDR}) with the relations between Tarasov-Bethe vectors and 
fusion-Bethe vectors (\ref{relationDR}), (\ref{drelationDR}), 
it is 
now straightforward to compute the scalar 
product between the off-shell state $\langle\Psi_m(u_1,\dots,u_m)|$ 
(\ref{dTBVdouble}) and the on-shell state 
$|\Psi_m(v_1,\dots,v_m)\rangle$ (\ref{TBVdouble}) 
\be\label{SLAVDRZF}
&&\langle\Psi_m(u_1,\dots,u_m)|\Psi_m(v_1,\dots,v_m)\rangle=\cosh^m(\eta)\, \bar s(u_1,\dots,u_m)\, \bar s(v_1,\dots,v_m)\non\\
&&\quad\quad\times\prod_{i=1}^m\frac{\delta_2\left(u_i+\frac{\eta}{2}\right)}{\sinh\left(2\left(v_i+\frac{3\eta}{2}\right)\right)\sinh\left(u_i+\frac{\eta}{2}-\xi^+\right)}
\prod_{j<i}^m\frac{\sinh(u_i+u_j+3\eta)}{\sinh(u_i+u_j+\eta)}
\non\\
&&\quad\quad\times
\frac{\textrm{det}_m\left(\frac{\partial}{\partial u_i}\delta\left(v_j+\frac{\eta}{2},u_1+\frac{\eta}{2},\dots,u_m+\frac{\eta}{2}\right))\right)}
{\textrm{det}_m\left(\bar\mF(v_i,u_j)\right)}\,.
\ee
We can rewrite (\ref{SLAVDRZF}) 
like (\ref{SLAVBEDR})
\be\label{SLAVBEDRZF}
&&\langle\Psi_m(u_1,\dots,u_m)|\Psi_m(v_1,\dots,v_m)\rangle\non\\&&\quad=
\prod_{j<k}^m\frac{\sinh(u_j+u_k)\sinh(u_j-u_k-2\eta)}{\sinh(u_j+u_k+\eta)\sinh(u_j-u_k-\eta)}\frac{\sinh(v_j+v_k)\sinh(v_j-v_k-2\eta)}{\sinh(v_j+v_k+\eta)\sinh(v_j-v_k-\eta)}
\non\\&&\quad\times
\prod_{i=1}^m
\frac{\sinh(2u_i+2\eta)\Delta_2(u_i)}{\sinh(2v_i+2\eta)\sinh\left(u_i+\frac{\eta}{2}-\xi^{+}\right)}\prod_{j<i}^m\frac{\sinh(u_i+u_j+3\eta)}{\sinh(u_i+u_j+\eta)}
\non\\&&\quad\times
\frac{\textrm{det}_m\left(\sinh(2\eta)\bar\mF(u_i,v_j)\bar\mF(v_j,u_i)\frac{\bar \mE_i(v_j)}{\bar\mQ_i(v_j)}\right)}
{\textrm{det}_m\left(\bar\mF(v_i,u_j)\right)}\,,
\ee
where
\be
\bar\mF(u,v)=\frac{1}{\sinh(u-v)\sinh(u+v+2\eta)}\,,
\ee
\textit{i.e.}, in terms of the Bethe ansatz (\ref{BPdoubleT}) and the 
Baxter polynomial (\ref{BaxterQbar}).

\section{Conclusion}\label{sec:conclusion}

We have considered the
Bethe vectors of the ZF model from two different perspectives, namely,
the fusion technique and Tarasov's
construction. For the single-row transfer matrix, the relations
between the two sets of Bethe vectors are
given by (\ref{relationSR},\ref{drelationSR}).  For the double-row
transfer matrix, the associated formulas are given by
(\ref{relationDR},\ref{drelationDR}).  By means of 
these simple
relations, we have computed the Slavnov scalar products for the Bethe
vectors within Tarasov's construction, see (\ref{SLAVBESRZF}) for
the single-row case and (\ref{SLAVBEDRZF}) for the double-row case.
The Slavnov formulas have been written in terms of Bethe ansatz and
Baxter polynomials, which seems to have more universal structure. 

We hope that our results can be useful in the study of the Slavnov
scalar products for other 19-vertex models, since they have similar
commutation relations in the 3-dimensional auxiliary space
representation.  One first step would be to derive the
formulas (\ref{SLAVBESRZF}) and (\ref{SLAVBEDRZF}) for the ZF model
without resorting to the fusion-Bethe vectors.  A particularly
important 19-vertex model is the Izergin-Korepin (IK)
model.  We hope that there exist scalar product formulas for IK
analogous to (\ref{SLAVBESRZF}) or (\ref{SLAVBEDRZF}), at least for
the quantum-group-invariant \cite{Nepomechie:1999jz} or root of unity
cases \cite{2014arXiv1411.2903G}, even though IK (unlike ZF) cannot be
obtained from some more elementary model by fusion.

\section*{Acknowledgments}
RP thanks Alexander Garbali for insightful discussions about the
scalar product of 19-vertex models.  This work was supported by the
S\~ao Paulo Research Foundation (FAPESP) and the University of Miami
under the SPRINT grant \#2016/50023-5.  Additional support was
provided by the Brazilian Research Council (CNPq) grant \#
310625/2013-0 and FAPESP \#2011/18729-1 (ALS), by a Cooper fellowship
and a Fulbright Specialist grant (RN), and by FAPESP/CAPES grant \# 2017/02987-8 (RP).  RN acknowledges
the hospitality at UFSCar, ICTP-SAIFR and AIMS-SA. RP thanks the University
of Miami for its warm hospitality.

\appendix

\section{Auxiliary formulas}\label{sec:auxiliary}
The auxiliary functions entering the off-shell equations (\ref{offshellTBVDR}) and (\ref{doffshellTBVDR}) are given by
\be
&&\mathcal{G}_1(u,u_j)=\frac{\cosh \left(u+\frac{\eta }{2}\right) \sinh (2 \eta ) \sinh (2
   (u+\eta )) }{4 \sinh
   \left(u-u_j\right) \sinh \left(u-\eta -u_j\right) \sinh \left(2
   \left(\eta +u_j\right)\right) \sinh \left(u+\eta +u_j\right) \sinh
   \left(u+2 \eta +u_j\right)}\non\\
&&\times 
\left(\sinh \left(u-\frac{\eta }{2}-2 u_j-\xi ^+\right)+\sinh \left(u+\frac{\eta
   }{2}+\xi ^+\right)-\sinh \left(3 u+\frac{3 \eta }{2}-\xi
   ^+\right)\right.\non\\&&\left.\quad\quad+\sinh \left(u+\frac{5 \eta }{2}+\xi ^+\right)+\sinh
   \left(u+\frac{7 \eta }{2}+2 u_j-\xi ^+\right)-\sinh
   \left(u+\frac{9 \eta }{2}+\xi ^+\right)\right)\,,
\ee
\be
\mathcal{G}_2(u,u_j)=\frac{2 \cosh \left(u+\frac{\eta }{2}\right) \sinh (\eta ) \sinh (2
   (u+\eta )) \sinh \left(u+\frac{3 \eta }{2}+\xi ^+\right)}{\sinh
   \left(u-\eta -u_j\right) \sinh \left(2 \left(\eta +u_j\right)\right)
   \sinh \left(u+\eta +u_j\right)}\,,
\ee
\be
\beta_1(u,u_j,u_k)&=&\sinh ^2(\eta ) \cosh (\eta )
\left(-\cosh \left(u-\frac{\eta }{2}\right)-\cosh \left(u+\frac{3 \eta
   }{2}\right)+\cosh \left(3 u+\frac{9 \eta }{2}\right)\right. \non\\&+& \left.\cosh \left(5
   u+\frac{11 \eta }{2}\right)\right)\frac{W_1(u,u_j,u_k)}{W_2(u,u_j,u_k)}\,
\ee
where
\be
&&W_1(u,u_j,u_k)=-\cosh \left(2 u +u_j-u_k-2 \xi ^+-3 \eta\right)+\cosh \left(3 u_j-u_k-2 \xi ^++\eta\right)\non\\&&-\cosh
   \left(2 u -u_j+u_k-2 \xi ^+-3 \eta\right)+\cosh \left(u_j+u_k-2 \xi ^++\eta\right)\non\\&&
   +\cosh \left(4 u+u_j+u_k-2 \xi ^++\eta\right)
      -\cosh \left(u_j+u_k-2 \xi ^++3 \eta\right)\non\\&&
      -\cosh \left(2 u +3u_j+u_k-2 \xi ^++3 \eta\right)+
      \cosh \left(3 u_k -u_j-2 \xi ^++\eta\right)
   \non\\&&-\cosh \left(2 u +u_j+3u_k-2 \xi ^++3 \eta\right)+
   \cosh \left(3 u_j+3 u_k-2 \xi ^++5 \eta\right)
   \non\\&&
   -\cosh \left(2 u -u_j-3 u_k+2\xi ^+-\eta\right)-\cosh \left(2 u -3 u_j-u_k+2 \xi ^+-\eta\right)
   \non\\&&
   +\cosh \left( -u_j-u_k+2 \xi^++\eta\right)+\cosh \left( -u_j-u_k+2 \xi ^++3 \eta\right)
   \non\\&&
   +
\cosh \left(4u -u_j-u_k+2 \xi ^++3 \eta\right)-\cosh \left(2 u +u_j-u_k+2 \xi ^++5 \eta\right)
   \non\\&&
   -\cosh \left(3 u_j-u_k+2 \xi^++3 \eta\right)+\cosh \left(3 u_j-u_k+2 \xi ^++5 \eta\right)
   \non\\&&
   -\cosh \left(2u -u_j+u_k+2 \xi ^++5 \eta\right)+\cosh \left(u_j+u_k+2 \xi ^++\eta\right)\non\\&&-\cosh \left(u_j+u_k+2 \xi ^++3 \eta\right)+\cosh\left(u_j+u_k+2 \xi ^++5 \eta\right)\non\\&&
   -\cosh \left(-u_j+3 u_k+2 \xi ^++3 \eta\right)+\cosh \left(-u_j+3 u_k+2 \xi ^++5 \eta\right)
   \non\\&&
   +\cosh \left(2 u-4 \eta -u_j-u_k\right)+\cosh \left(2u -u_j-u_k-2 \eta\right)\non\\&&
   -\cosh \left(2 (u+\eta)-u_j-u_k\right)-\cosh \left(4 u +u_j-u_k+2 \eta\right)\non\\&&+
\cosh \left(2 (u+\eta)+3u_j-u_k\right)-\cosh \left(4 u -u_j+u_k+2 \eta\right)
   \non\\&&
   +
\cosh \left(2u +u_j+u_k+4 \eta\right)+\cosh \left(2 u +u_j+u_k+6 \eta\right)
   \non\\&&
   +\cosh \left(2 (u+\eta)+u_j+u_k\right)-\cosh \left(3 u_j+u_k+6\eta\right)\non\\&&
   +\cosh \left(2 (u+\eta)-u_j+3 u_k\right)-\cosh \left(u_j+3u_k+6 \eta\right)+
   \cosh \left(2 u+u_j-3 u_k\right)\non\\&&+\cosh \left(2 u-u_j-u_k\right)-3 \cosh\left(u_j-u_k\right)-\cosh \left(3 \left(u_j-u_k\right)\right)\non\\&&+\cosh \left(2 u-3 u_j+u_k\right)-\cosh
   \left(2 u+u_j+u_k\right)\,,
\ee
\be
&&W_2(u,u_j,u_k)=32 \sinh \left(u-u_j\right) \sinh \left(u-u_k\right)
\sinh \left(u -u_j-\eta\right) \sinh \left(2\left(u_j+\eta\right)\right)\non\\&&\times
\sinh \left(u +u_j+\eta\right) \sinh \left(u +u_j+2 \eta\right) \sinh\left( u-u_k-\eta\right) \sinh \left(2 \left(u_k+\eta\right)\right) \non\\&&\times
   \sinh \left(u+u_k+\eta\right)
   \sinh \left(u+u_k+2 \eta\right) \sinh \left(u_j-u_k+\eta\right)
   \sinh \left( u_k-u_j+\eta\right)
   \sinh \left(u_j+u_k+\eta\right) \non\\&&\times\sinh \left(u_j+u_k+2 \eta\right) \sinh \left(u_j-\xi ^++\frac{\eta}{2}\right) \sinh \left(u_k-\xi ^++\frac{\eta }{2}\right)\,,
\ee
\be
&&\beta_2(u,u_j,u_k)=\frac{2 \cosh \left(u+\frac{\eta }{2}\right) \sinh ^2(\eta ) \sinh (2 \eta ) \sinh (2 (u+\eta )) \sinh (2
   u+3 \eta )}{\sinh \left(u-u_j\right) \sinh \left(u-u_j-\eta\right) \sinh \left(u+u_j+\eta\right)
   \sinh \left(u +u_j+2 \eta\right)}\non\\&&\times
   \frac{\sinh \left(2 \left(u_k+\eta \right)\right)}
   {\sinh \left(2 \left(u_j+\eta\right)\right) \sinh \left(u_j-u_k\right) \sinh \left(u_j-u_k+\eta \right)  \sinh \left(u_j+u_k+\eta\right) \sinh \left(u_j+u_k+2 \eta\right)}
   \non\\&&\times
   \frac{\sinh \left(u_j-\frac{\eta }{2}-\xi^{+}\right) \sinh \left(u_j+\frac{5 \eta }{2}+\xi^{+}\right)}{\sinh \left(u_k+\frac{\eta}{2}-\xi^{+}\right)}\,.
\ee


\begin{thebibliography}{10}

\bibitem{Zamolodchikov:1980ku}
A.~B. Zamolodchikov and V.~A. Fateev, ``{Model Factorized S Matrix And An
  Integrable Heisenberg Chain With Spin 1.},'' {\em Sov. J. Nucl.
  Phys.} {\bfseries 32} (1980) 298--303.
[Yad. Fiz.32,581(1980)].

\bibitem{Faddeev:1979gh}
L.~D. Faddeev, E.~K. Sklyanin, and L.~A. Takhtajan, ``{The Quantum Inverse
  Problem Method. 1},'' {\em Theor. Math. Phys.} {\bfseries 40} (1980)
  688--706.
[Teor. Mat. Fiz.40,194(1979)].

\bibitem{Faddeev:1996iy}
L.~D. Faddeev, ``{How algebraic Bethe ansatz works for integrable models},'' in
  {\em Sym\'etries Quantiques (Les Houches Summer School Proceedings vol 64)},
  A.~Connes, K.~Gawedzki, and J.~Zinn-Justin, eds., pp.~149--219.
\newblock North Holland, 1998.
\newblock
\href{http://arxiv.org/abs/hep-th/9605187}{{\ttfamily arXiv:hep-th/9605187
  [hep-th]}}.
\newblock

\bibitem{Sklyanin:1988yz}
E.~K. Sklyanin, ``{Boundary Conditions for Integrable Quantum Systems},''
\href{http://dx.doi.org/10.1088/0305-4470/21/10/015}{{\em J. Phys.} {\bfseries
  A21} (1988) 2375--289}.

\bibitem{Kulish:1981gi}
P.~P. Kulish, N.~{\relax Yu}. Reshetikhin, and E.~K. Sklyanin, ``{Yang-Baxter
  Equation and Representation Theory. 1.},''
\href{http://dx.doi.org/10.1007/BF02285311}{{\em Lett. Math. Phys.} {\bfseries
  5} (1981) 393--403}.

\bibitem{Kulish:1981bi}
P.~P. Kulish and E.~K. Sklyanin, ``{Quantum spectral transform method. Recent
  developments},''
{\em Lect. Notes Phys.} {\bfseries 151} (1982) 61--119.

\bibitem{Babujian:1983ae}
H.~M. Babujian, ``{Exact Solution Of The Isotropic Heisenberg Chain With
  Arbitrary Spins: Thermodynamics Of The Model },''
\href{http://dx.doi.org/10.1016/0550-3213(83)90668-5}{{\em Nucl. Phys.}
  {\bfseries B215} (1983) 317--336}.

\bibitem{A.:1982zz}
L.~A. Takhtajan, ``{The picture of low-lying excitations in the isotropic
  Heisenberg chain of arbitrary spins},''
\href{http://dx.doi.org/10.1016/0375-9601(82)90764-2}{{\em Phys. Lett.}
  {\bfseries A87} (1982) 479--482}.

\bibitem{Kulish:1983md}
P.~P. Kulish and N.~{\relax Yu}. Reshetikhin, ``{Quantum linear problem for the
  Sine-Gordon equation and higher representation},''
  \href{http://dx.doi.org/10.1007/BF01084171}{{\em J. Sov. Math.} {\bfseries
  23} (1983) 2435--2441}.
[Zap. Nauchn. Semin.101,101(1981)].

\bibitem{SOGO198451}
K.~Sogo, ``{Ground state and low-lying excitations in the Heisenberg XXZ chain
  of arbitrary spin S},''
  \href{http://dx.doi.org/https://doi.org/10.1016/0375-9601(84)90588-7}{{\em
  Physics Letters A} {\bfseries 104} no.~1, (1984) 51 -- 54}.

\bibitem{Babujian:1986ua}
H.~M. Babujian and A.~M. Tsvelik, ``{Heisenberg magnet with an arbitrary spin
  and anisotropic chiral field},''
\href{http://dx.doi.org/10.1016/0550-3213(86)90405-0}{{\em Nucl. Phys.}
  {\bfseries B265} (1986) 24--44}.

\bibitem{Kirillov1986}
A.~N. Kirillov and N.~Y. Reshetikhin, ``{Exact solution of the Heisenberg XXZ
  model of spin s},'' \href{http://dx.doi.org/10.1007/BF01083768}{{\em Journal
  of Soviet Mathematics} {\bfseries 35} no.~4, (1986) 2627--2643}.

\bibitem{Gohmann:2010se}
F.~Gohmann, A.~Seel, and J.~Suzuki, ``{Correlation functions of the integrable
  isotropic spin-1 chain at finite temperature},''
  \href{http://dx.doi.org/10.1088/1742-5468/2010/11/P11011}{{\em J. Stat.
  Mech.} {\bfseries 1011} (2010) P11011},
\href{http://arxiv.org/abs/1008.4440}{{\ttfamily arXiv:1008.4440
  [cond-mat.str-el]}}.

\bibitem{Beisert:2015msa}
N.~Beisert, M.~de~Leeuw, and P.~Nag, ``{Fusion for the one-dimensional Hubbard
  model},'' \href{http://dx.doi.org/10.1088/1751-8113/48/32/324002}{{\em J.
  Phys.} {\bfseries A48} no.~32, (2015) 324002},
\href{http://arxiv.org/abs/1503.04838}{{\ttfamily arXiv:1503.04838 [math-ph]}}.

\bibitem{Mezincescu:1990fc}
L.~Mezincescu, R.~I. Nepomechie, and V.~Rittenberg, ``{Bethe Ansatz Solution of
  the Fateev-Zamolodchikov Quantum Spin Chain With Boundary Terms},''
\href{http://dx.doi.org/10.1016/0375-9601(90)90016-H}{{\em Phys. Lett.}
  {\bfseries A147} (1990) 70--78}.

\bibitem{Mezincescu:1991ke}
L.~Mezincescu and R.~I. Nepomechie, ``{Fusion procedure for open chains},''
{\em J. Phys.} {\bfseries A25} (1992) 2533--2544.

\bibitem{Nepomechie:2015zwa}
R.~I. Nepomechie and R.~A. Pimenta, ``{Fusion for AdS/CFT boundary
  S-matrices},'' \href{http://dx.doi.org/10.1007/JHEP11(2015)161}{{\em JHEP}
  {\bfseries 11} (2015) 161},
\href{http://arxiv.org/abs/1509.02426}{{\ttfamily arXiv:1509.02426 [hep-th]}}.

\bibitem{Tarasov}
V.~Tarasov, ``{Algebraic Bethe ansatz for the Izergin-Korepin R matrix},''
  \href{http://dx.doi.org/10.1007/BF01028578}{{\em Theor. Math. Phys.}
  {\bfseries 76} no.~2, (1988) 793--803}.

\bibitem{izergin1981}
A.~G. Izergin and V.~E. Korepin, ``{The inverse scattering method approach to
  the quantum Shabat-Mikhailov model},'' {\em Comm. Math. Phys.} {\bfseries 79}
  no.~3, (1981) 303--316.

\bibitem{Jones:2017}
V.~F. Jones, ``{Scale invariant transfer matrices and Hamiltonians},''
{\em J. Phys.} {\bfseries A51} (2018) 104001,
  \href{http://arxiv.org/abs/1706.00515}{{\ttfamily arXiv:1706.00515
  [math.OA]}}.

\bibitem{Fan:1997wpk}
H.~Fan, ``{Bethe ansatz for the Izergin-Korepin model},''
\href{http://dx.doi.org/10.1016/S0550-3213(97)00023-0}{{\em Nucl. Phys.}
  {\bfseries B488} (1997) 409--425}.

\bibitem{LimaSantos:1998te}
A.~Lima-Santos, ``{Bethe ansatz for nineteen vertex models},''
  \href{http://dx.doi.org/10.1088/0305-4470/32/10/004}{{\em J. Phys.}
  {\bfseries A32} (1999) 1819--1839},
\href{http://arxiv.org/abs/hep-th/9807219}{{\ttfamily arXiv:hep-th/9807219
  [hep-th]}}.

\bibitem{Kurak:2004ip}
V.~Kurak and A.~Lima-Santos, ``{Algebraic Bethe Ansatz for the
  Zamolodchikov-Fateev and Izergin-Korepin models with open boundary
  conditions},'' \href{http://dx.doi.org/10.1016/j.nuclphysb.2004.08.040}{{\em
  Nucl. Phys.} {\bfseries B699} (2004) 595--631},
\href{http://arxiv.org/abs/nlin/0406050}{{\ttfamily arXiv:nlin/0406050
  [nlin-si]}}.

\bibitem{Sla89}
N.~A. Slavnov, ``{Calculation of scalar products of wave functions and form
  factors in the framework of the algebraic Bethe ansatz},''
  \href{http://dx.doi.org/10.1007/BF01016531}{{\em Theoret. and Math. Phys.}
  {\bfseries 79} (1989) 502--508}.

\bibitem{Kitanine:2001bq}
N.~Kitanine, ``{Correlation functions of the higher spin XXX chains},''
  \href{http://dx.doi.org/10.1088/0305-4470/34/39/314}{{\em J. Phys.}
  {\bfseries A34} (2001) 8151},
\href{http://arxiv.org/abs/math-ph/0104016}{{\ttfamily arXiv:math-ph/0104016
  [math-ph]}}.

\bibitem{Deguchi:2009zz}
T.~Deguchi and C.~Matsui, ``{Form factors of integrable higher-spin XXZ chains
  and the affine quantum-group symmetry},''
  \href{http://dx.doi.org/10.1016/j.nuclphysb.2009.01.002,
  10.1016/j.nuclphysb.2011.05.013}{{\em Nucl. Phys.} {\bfseries B814} (2009)
  405--438}, \href{http://arxiv.org/abs/0807.1847}{{\ttfamily arXiv:0807.1847
  [cond-mat.stat-mech]}}.
[Erratum: Nucl. Phys.B851,238(2011)].

\bibitem{Wang2002633}
Y.-S. Wang, ``The scalar products and the norm of {B}ethe eigenstates for the
  boundary {XXX} {H}eisenberg spin-1/2 finite chain,''
  \href{http://dx.doi.org/http://dx.doi.org/10.1016/S0550-3213(01)00610-1}{{\em
  Nucl.Phys.} {\bfseries B622} no.~3, (2002) 633 -- 649}.

\bibitem{Belliard:2016ist}
S.~Belliard and R.~A. Pimenta, ``{Slavnov and Gaudin-Korepin formulas for
  models without U (1) symmetry: the XXX chain on the segment},''
\href{http://dx.doi.org/10.1088/1751-8113/49/17/17LT01}{{\em J. Phys.}
  {\bfseries A49} no.~17, (2016) 17LT01}.

\bibitem{Kitanine:2007bi}
N.~Kitanine, K.~K. Kozlowski, J.~M. Maillet, G.~Niccoli, N.~A. Slavnov, and
  V.~Terras, ``{Correlation functions of the open XXZ chain I},''
  \href{http://dx.doi.org/10.1088/1742-5468/2007/10/P10009}{{\em J. Stat.
  Mech.} {\bfseries 0710} (2007) P10009},
\href{http://arxiv.org/abs/0707.1995}{{\ttfamily arXiv:0707.1995 [hep-th]}}.

\bibitem{Kitanine:1999rfm}
N.~Kitanine, J.~M. Maillet, and V.~Terras, ``{Form factors of the XXZ
  Heisenberg spin-$\frac 1 2$ finite chain},''
\href{http://dx.doi.org/10.1016/S0550-3213(99)00295-3}{{\em Nucl. Phys.}
  {\bfseries B554} (1999) 647--678}.

\bibitem{Nepomechie:1999jz}
R.~I. Nepomechie, ``{Nonstandard coproducts and the Izergin-Korepin open spin
  chain},'' \href{http://dx.doi.org/10.1088/0305-4470/33/2/101}{{\em J. Phys.}
  {\bfseries A33} (2000) L21--L26},
\href{http://arxiv.org/abs/hep-th/9911232}{{\ttfamily arXiv:hep-th/9911232
  [hep-th]}}.

\bibitem{2014arXiv1411.2903G}
A.~{Garbali}, ``{The domain wall partition function for the Izergin-Korepin
  19-vertex model at a root of unity},'' {\em ArXiv e-prints} (Nov., 2014) ,
  \href{http://arxiv.org/abs/1411.2903}{{\ttfamily arXiv:1411.2903 [math-ph]}}.

\end{thebibliography}

\providecommand{\href}[2]{#2}\begingroup\raggedright\endgroup

\end{document}